\documentclass{aa}
\usepackage[varg]{txfonts}
\usepackage{graphicx}

\usepackage[dvipsnames]{xcolor}
\usepackage[colorlinks,citecolor=RoyalBlue,linkcolor=blue]{hyperref}

\usepackage{etoolbox}
\makeatletter
\patchcmd\@combinedblfloats{\box\@outputbox}{\unvbox\@outputbox}{}{%
   \errmessage{\noexpand\@combinedblfloats could not be patched}%
}%
 \makeatother

\begin{document}

\title{Confirmation of double peaked Ly$\alpha$ emission at $z=6.593$ \thanks{Based on observations obtained with the Very Large Telescope, programs: 294.A-5039, 099.A-0254 \& 100.A-0213}}
\subtitle{Witnessing a galaxy directly contributing to the reionisation of the Universe}

\author{Jorryt Matthee\inst{1}\thanks{e-mail: matthee@strw.leidenuniv.nl}
  \and David Sobral\inst{2} 
     \and Max Gronke \inst{3}
     \and Ana Paulino-Afonso \inst{2,4,5}
     \and Mauro Stefanon \inst{1}
     \and Huub R\"ottgering \inst{1}} 


\institute{Leiden Observatory, Leiden University, PO\ Box 9513, NL-2300 RA, Leiden, The Netherlands
  \and Department of Physics, Lancaster University, Lancaster, LA1 4YB, UK
    \and Department of Physics, University of California, Santa Barbara, CA 93106, USA
             \and Instituto de Astrof\'isica e Ci\^encias do Espaco, Universidade de Lisboa, OAL, Tapada da Ajuda, 1349-018 Lisboa, Portugal
           \and Departamento de F\'isica, Faculdade de Ci\^encias, Universidade de Lisboa, Edif\'icio C8, Campo
Grande, 1749-016 Lisboa, Portugal  
             } 


\abstract{Distant luminous Lyman-$\alpha$ emitters (LAEs) are excellent targets for spectroscopic observations of galaxies in the epoch of reionisation (EoR). We present deep high-resolution ($R=5000$) VLT/X-SHOOTER observations, along with an extensive collection of photometric data of `COLA1', a proposed double peaked LAE at $z=6.6$ \citep{Hu2016}. We rule out that COLA1's emission line is an [O{\sc ii}] doublet at $z=1.475$ on the basis of i) the asymmetric red line-profile and flux ratio of the peaks (blue/red=$0.31\pm0.03$) and ii) an unphysical [O{\sc ii}]/H$\alpha$ ratio ([O{\sc ii}]/H$\alpha > 22$). We show that COLA1's observed $B$-band flux is explained by a faint extended foreground LAE, for which we detect Ly$\alpha$ and [O{\sc iii}] at $z=2.142$. We thus conclude that COLA1 is a real double-peaked LAE at $z=6.593$, the first discovered at $z>6$, confirming the result from \cite{Hu2016}. COLA1 is UV luminous (M$_{1500}=-21.6\pm0.3$), has a high equivalent width (EW$_{0, \rm Ly\alpha}=120^{+50}_{-40}$ {\AA}) and very compact Ly$\alpha$ emission ($r_{50,\rm Ly\alpha} = 0.33^{+0.07}_{-0.04}$ kpc). Relatively weak inferred H$\beta$+[O{\sc iii}] line-emission from {\it Spitzer}/IRAC indicates an extremely low metallicity of $Z<1/20\,Z_{\odot}$ or reduced strength of nebular lines due to high escape of ionising photons. The small Ly$\alpha$ peak separation of $220\pm20$ km s$^{-1}$ implies a low H{\sc i} column density and an ionising photon escape fraction of $\approx15-30$ \%, providing the first direct evidence that such galaxies contribute actively to the reionisation of the Universe at $z>6$. Based on simple estimates, we find that COLA1 could have provided just enough photons to reionise its own $\approx0.3$ pMpc (2.3 cMpc) bubble, allowing the blue Ly$\alpha$ line to be observed. However, we also discuss alternative scenarios explaining the detected double peaked nature of COLA1. Our results show that future high-resolution observations of statistical samples of double peaked LAEs at $z>5$ are a promising probe of the occurrence of ionised regions around galaxies in the EoR.}

\maketitle 

\section{Introduction} \label{sec:intro}
The epoch of reionisation (EoR) is the last phase transition of the Universe. It occurred when the neutral hydrogen of the intergalactic medium (IGM) became reionised \citep[e.g.][]{Madau2017}. In spite of the increasingly precise measurements of the global progress of reionisation and its patchiness \citep[e.g.][]{Becker2015,Davies2017,Banados2018,Bosman2018}, its evolution and drivers are still largely unknown. One of the main probes of reionisation is the Lyman-$\alpha$ (Ly$\alpha$) line (see \citealt{Dijkstra2007} for a review). Due to the low Ly$\alpha$ transmission in a partially neutral IGM, the progress of reionisation can be mapped out using the strength of detected Ly$\alpha$ emission. This is usually done through the equivalent width distribution among high-redshift galaxies \citep[e.g.][]{Pentericci2014,Mason2018}, through the evolution in the luminosity function \citep[e.g.][]{Zheng2017} and/or its clustering signal \citep[e.g.][]{Jensen2014,Kakiichi2018}.

Recently, wide-field narrow-band surveys have been highly efficient in identifying and confirming luminous Lyman-$\alpha$ emitters (LAEs) into the EoR ($z\approx7$; e.g. \citealt{Matthee2015,Hu2016,Santos2016,Bagley2017,Zheng2017,Shibuya2017}). The number densities of extremely luminous LAEs at $z\sim7$ (L$_{\rm Ly\alpha}>2\times10^{43}$ erg s$^{-1}$) are higher \citep{Matthee2015,Bagley2017} than expected based on older, smaller surveys \citep[e.g.][]{Ouchi2010}. While the number densities of faint LAEs decrease by 0.5 dex at $z>6$ \citep{Ouchi2010}, the number densities of the most luminous sources are relatively constant between $z\approx5-7$ \citep[e.g.][]{Santos2016,Zheng2017}, or decrease only slightly by $\approx0.2-0.3$ dex \citep[e.g.][]{Konno2018}. These luminous sources likely reside in early ionised bubbles \citep{Hu2016,Stark2017,Mason2018Letter}, facilitating their Ly$\alpha$ observability during the EoR \citep[e.g.][]{Matthee2015,Weinberger2018,Songaila2018}.

Due to their luminosity, luminous LAEs are ideal for spectroscopic follow-up \citep[e.g.][]{Sobral2015,Hu2016,Matthee2017spec,Shibuya2017}. The most Ly$\alpha$-luminous example among these sources at $z\approx6.6$ is CR7 \citep{Sobral2015,Sobral2017}. CR7 has spectacular properties: strong, narrow Ly$\alpha$ emission (EW$_0 \approx 210$ {\AA}), intense H$\beta$ and/or [O{\sc iii}] emission \citep{Bowler2017} and there are indications of strong high-ionisation rest-frame UV emission lines such as He{\sc ii} \citep{Sobral2017}. High resolution {\it HST} and ALMA observations reveal multiple components in the rest-frame UV separated by $\sim5$kpc and spatially and spectrally resolved [C{\sc ii}] FIR cooling-line emission \citep{Matthee2017ALMA}, indicating the build-up of a central galaxy through accretion of satellite galaxies. 
  
Recently, \cite{Hu2016} presented the spectroscopic confirmation of a potentially even more luminous LAE at $z=6.593$, `COLA1', with an unexpected double peaked Ly$\alpha$ line-profile. The presence of multiple peaks in COLA1 had been unprecedented in $z>6$ LAEs \citep[e.g.][]{Matthee2017spec,Shibuya2017}, but \cite{Songaila2018} show that roughly a third of a sample of extremely luminous LAEs at $z=6.6$ may have such complex line shapes. The transmission on the blue part of the Ly$\alpha$ line is typically very low \citep[e.g.][]{Dijkstra2007,Laursen2011}, particularly due to residual neutral hydrogen in the halos of galaxies \citep[e.g.][]{Mesinger2015} and the increasing neutral fraction of the inter galactic medium \citep[e.g.][]{Hu2010}. 
 
Several effects help to increase the observability of Ly$\alpha$ at high redshift: (i) the redshift of the Ly$\alpha$ line with respect to the surrounding gas (e.g. due to outflows), and (ii) the Hubble expansion that may redshift the Ly$\alpha$ line out of the resonance frequency of the neutral gas, thus lowering the optical depth in case the galaxy is surrounded by a large, highly ionised region. 

However, due to the large-scale infall of gas which shifts the absorption $\sim100$ km s$^{-1}$ towards the red, combined with the extended wing of the absorption profile of the highly neutral IGM, simulations and analytical models expect that even the red part of the Ly$\alpha$ emission is diminished significantly \citep[e.g.][]{Haiman2002,Mesinger2004,Dijkstra2007,Laursen2011,Mesinger2015,Sadoun2017,Weinberger2018}. Furthermore, radiative transfer processes in the interstellar medium (ISM) even shift the blue peak to lower velocities; e.g. \citealt{Neufeld1990}. Therefore, a detection of intrinsically blue-shifted Ly$\alpha$ emission above $z\gtrsim6$ is theoretically unexpected \citep[e.g.][]{Mason2018}.

Since the transmission of the blue part of the Ly$\alpha$ line depends directly on the ionisation state of the environment of the emitting galaxy, COLA1 could open up a completely new probe of reionisation, if the detection of the blue Ly$\alpha$ line is confirmed. With a larger sample, the evolution of the Ly$\alpha$ line asymmetry (i.e., comparing the blue and red fluxes) can be quantified more robustly, adding an interesting additional observable to the evolution of LAE clustering and EW. Interestingly, the Ly$\alpha$ peak separation correlates strongly with the escape of Lyman-Continuum (LyC) radiation \citep[e.g.][]{Verhamme2015,Vanzella2018,Izotov2018}. Therefore, COLA1 could also provide the first direct evidence that galaxies actively contribute to the reionisation of the Universe. 

However, the double peaked line could also mean that COLA1 is an [O{\sc ii}]$_{3727,3730}$ emitter at $z=1.475$ (see \citealt{Songaila2018} for a detailed investigation on the fraction of such interlopers). We therefore test whether the line may have been misinterpreted, in which case COLA1 is a low-redshift interloper, and not a LAE at $z=6.6$. Because of the unique potential of the COLA1 galaxy, we investigate its nature independently using our own narrow-band data (from \citealt{Sobral2013}, not used in \citealt{Matthee2015}), new high spectral resolution VLT/X-SHOOTER observations covering a wavelength range from $0.3$ to $2.5\mu$m and the wealth of public available data from the COSMOS survey which were not explored in \cite{Hu2016}. 

In this work, we first summarise the available photometric information and present the new X-SHOOTER observations in \S $\ref{sec:data}$. We analyse the spectrum in detail and conclude on the redshift of COLA1 in \S $\ref{sec:analysis}$. \S $\ref{sec:COLA1_properties}$ presents the properties of COLA1 that we measure from the spectroscopic and photometric data available. We discuss the implications in \S $\ref{sec:discussion}$ and summarise the results in \S $\ref{sec:summary}$. Throughout this work we adopt a flat $\Lambda$CDM cosmology with $\Omega_{\Lambda, 0}=0.7$, $\Omega_{\rm M, 0}=0.3$ and H$_{0}=70$ km s$^{-1}$ Mpc$^{-1}$, a \cite{Salpeter1995} initial mass function and magnitudes in the AB system \citep{Oke1974}.

\section{Data} \label{sec:data}
\subsection{Photometric measurements} 
COLA1 \citep{Hu2016} is located in the COSMOS field \citep{Capak2007,Scoville2007} and hence public imaging data are available in $\approx30$ filters from the UV to IR \citep[e.g.][]{Ilbert2009}. This includes high resolution data in the F814W filter from the {\it Hubble Space Telescope (HST)}/Advanced Camera for Surveys \citep{Koekemoer2007}. COLA1 is located in a shallow region from the UltraVISTA survey \citep{McCracken2012} and is covered by {\it Spitzer}/IRAC imaging from the SPLASH program \citep[e.g.][]{Steinhardt2014}\footnote{http://splash.caltech.edu}. We use optical data from CFHT/Megacam, Subaru/Suprime-cam \citep{Taniguchi2007}\footnote{Available through http://irsa.ipac.caltech.edu/data/COSMOS/} and near-infrared data from UltraVISTA DR3. COLA1 is furthermore covered in Subaru/Suprime-cam NB921 images from \cite{Sobral2012,Sobral2013}, similar to the NB921 imaging from Hyper Suprime-Cam from \cite{Hu2016}. We show thumbnail images in the $u$, $B$, $z'$, NB921, F814W, $Y$, $J$, $H$, [3.6] and [4.5] bands in Fig. $\ref{fig:thumbnails}$, where red contours illustrate the location of line-emission (measured from the NB921-$z$ image). We note that we have confirmed the astrometric alignment of the different images using the position of $\approx30$ stars and galaxies around COLA1. Our photometric measurements are summarised in Table $\ref{tab:photometry}$. The measurements on ground-based images in bands blue-wards of 900 nm are conducted with 1.2$''$ diameter apertures optimised to estimate the S/N of potential detections. Measurements in NB921, $z'$, $Y$, $J$ and $H$ are performed with 2$''$ diameter apertures.

\subsubsection{NB921 Narrow-band data}
Combining the Suprime-cam NB921 narrow-band data \citep{Sobral2013} with public $z'$ band data, we measure a total line-flux of $5.8^{+1.2}_{-1.1}\times10^{-17}$ erg s$^{-1}$ cm$^{-2}$. As the mean transmission of the NB921 filter is 69 \% at the wavelength where the emission-line is detected, we correct this line-flux to $8.2^{+1.7}_{-1.6}\times10^{-17}$ erg s$^{-1}$ cm$^{-2}$. Spreading this line-flux over the width of the $z'$ band filter results in $z=25.7$, meaning that the full $z'$ band flux can be explained by line-emission. We use the $Y$ band ($Y=25.2^{+0.4}_{-0.3}$, $\rm S/N\approx3$, measured in a 2$''$ diameter aperture and applying an aperture correction of $-0.3$ mag) to obtain a weak constraint on the observed equivalent width of EW$_{\rm obs} = 900^{+340}_{-240}$ {\AA} assuming a UV slope $\beta=-2.0$. The $Y$ band magnitude is in good agreement with the measurement in \cite{Hu2016} based on Subaru data. The line-flux we measure is a factor $\approx2$ lower than \cite{Hu2016}. We have checked that this is not due to the use of a smaller aperture (2$''$ used here versus 3$''$ diameter in \citealt{Hu2016}), as using a 3$''$ aperture would only increase the flux by 10 \%. We note that our photometry in 3$''$ apertures is consistent with \cite{Hu2016}, meaning that the difference is due to the method to calculate the line-flux.

\subsubsection{Ground-based optical data}
Visual inspection of the thumbnail images reveals potential detections in the $u$ and $B$ filters. While the potential flux in the $u$ band is offset, this is not as clear in the $B$ band. We measure the significance of the flux in the $B$ band by performing photometry centred on the NB921 detection in an optimised 1.2$''$ diameter aperture. We choose such small aperture to optimise the S/N ratio (the PSF-FWHM of this data is $0.7-0.8''$) and minimise contamination from nearby objects. \footnote{We note that we do not use measurements with different apertures to measure colours.} The noise level is measured from the standard deviation of 1000 empty aperture measurements located on random sky positions around COLA1. We measure flux at the 2.4 $\sigma$ level with $B=28.4^{+0.6}_{-0.4}$ (see Fig. $\ref{fig:thumbnails}$, where we use a high contrast to highlight potential detections). No flux above the 2$\sigma$ level is detected in the $u$, $V$, $R$ and $I$ filters using the same aperture centered on COLA1. We measure flux at the 2$\sigma$ level in the stacked $BVRI$ image. While the significance levels of these measurements is low, they may be troublesome for a galaxy at $z>6.5$, motivating the need for careful spectroscopic analysis.

\subsubsection{High resolution optical data} 
The {\it HST} imaging in the F814W filter (with $>20$ \% transmission between $\lambda=6988-9577$ {\AA}, $\lambda_{\rm eff} =7985$ {\AA}, PSF-FWHM $=0.095''$; \citealt{Koekemoer2007}) reveals a faint point-like detection at the position of COLA1 (Fig. $\ref{fig:thumbnails}$). Similarly to the $u$ and $B$ imaging, F814W imaging also clearly shows another source 1.2$''$ south-west of COLA (identified in the {\it HST} thumbnail in Fig. $\ref{fig:thumbnails}$). This object has ID 593625 in the \cite{Laigle2016} catalogue (with photometric redshift $p_z=1.9^{+0.2}_{-0.1}$) and we will refer to it with that ID from now on. COLA1 is detected at 3$\sigma$, with $\rm F814W = 26.6^{+0.4}_{-0.3}$ measured in a 0.6$''$ diameter aperture and corrected for aperture losses using tabulated encircled fluxes from \cite{Bohlin2016}. ID 593625 has a F814W magnitude of $\rm F814W = 26.5^{+0.3}_{-0.3}$. If continuum flux at this level contributes to the NB921 photometry it could contribute to the 10 \% increase in the line-flux of COLA1 measured with 3$''$ apertures.

\begin{table}
\centering
\caption{The coordinates and photometric measurements of COLA1. Magnitudes are in the AB system. Magnitude limits are at the 2$\sigma$ level. $^*$As the $H$ band photometry is contaminated by ID 593625 we only provide it as a lower limit. } \label{tab:photometry}
\begin{tabular}{lc} \hline
\bf Coordinates & \\ 
R.A. & 10:02:35.37 (J2000) \\
Dec. & +02:12:13.9 (J2000) \\ \hline
\bf Photometry & \\
$B$ &$28.4^{+0.6}_{-0.4}$ \\
$V$ &$>27.8$\\
$R$ &$>27.7$\\
$I$ & $>27.6$ \\
F814W & $26.6^{+0.4}_{-0.3}$ \\ 
NB921 & $23.6^{+0.1}_{-0.1}$\\
$z'$ & $25.5^{+0.5}_{-0.3}$\\
$Y$ & $25.2^{+0.4}_{-0.3}$\\
$J$ & $24.8^{+0.5}_{-0.3}$\\ 
$H$ & $>24.8^*$\\
$K_s$ & $>24.4$ \\
$[3.6]$ & $24.4^{+0.2}_{-0.1}$ \\
$[4.5]$ & $24.6^{+0.3}_{-0.2}$\\
\hline
\end{tabular}
\end{table}

\subsubsection{Photometric redshift and the {\it Spitzer}/IRAC view}
\begin{figure}
 \centering
	\includegraphics[width=8.2cm]{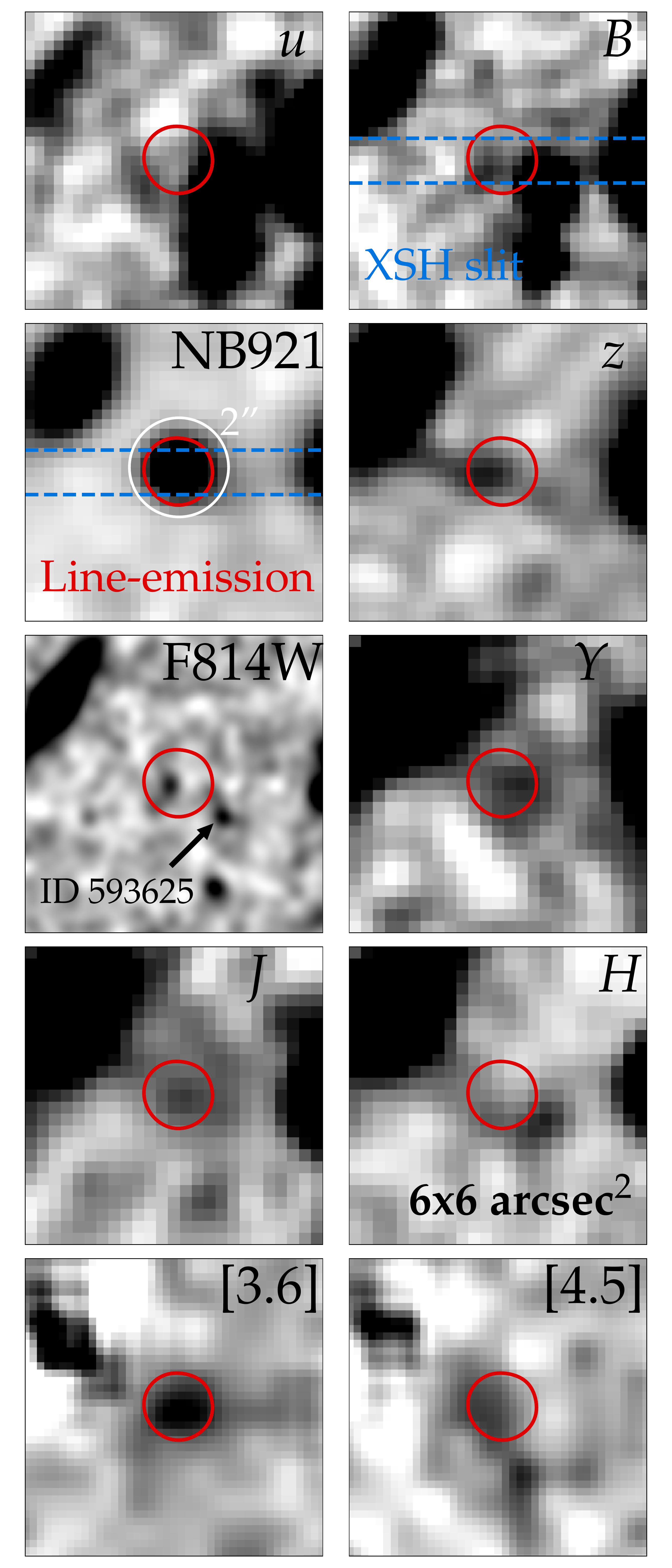} 
    \caption{High contrast thumbnail images of $6\times6$ arcsec$^2$ ($32\times32$ kpc$^2$) centered on the NB921 detection of COLA1. The red contour shows the 2$\sigma$ level of the NB921-$z'$ image (tracing line-emission). The white circle illustrates a 2$''$ aperture. X-SHOOTER observations were performed with a slit alignment PA=0. A positional shift upwards along the slit corresponds to a shift to the right (west) in these thumbnails. The IRAC [3.6] and [4.5] images have been cleaned using {\it HST}/F814W as a prior following \citet{Stefanon2017}.}
    \label{fig:thumbnails}
\end{figure}

COLA1 is present in the public COSMOS2015 catalogue (\citealt{Laigle2016}; ID 593751), where it has a photometric redshift of $p_z=0.99^{+0.12}_{-0.11}$. However, the photometry that is used to estimate this redshift is measured with 2$''$ diameter apertures and may suffer contamination from ID 593625. We re-measure the photometry in the {\it Spitzer}/IRAC $[3.6]$ and $[4.5]$ filters using SPLASH data. We follow the procedure as last described in \cite{Stefanon2017}, where the IRAC images are de-confused based on the {\it HST}/F814W images using the \texttt{mophongo} software \citep{Labbe2006,Labbe2015}. We measure $[3.6]=24.4^{+0.2}_{-0.1}$ and $[4.5]=24.6^{+0.3}_{-0.2}$ in 1.8$''$ diameter apertures and including an aperture correction as described in \cite{Labbe2015}, see Fig. $\ref{fig:thumbnails}$. This results in a moderately blue colour $[3.6]-[4.5]=-0.2\pm0.3$, although with significant uncertainties (see also \citealt{Harikane2018}, who measure $[3.6]-[4.5]=-0.2\pm0.1$).

\subsection{X-SHOOTER spectroscopic observations \& data reduction} 
We observed COLA1 with the X-SHOOTER echelle spectrograph on UT2 of the Very Large Telescope as part of ESO program 0100.A-0213 (PI Matthee). X-SHOOTER observes in three arms simultaneously: UVB ($\lambda=0.30-0.55\mu$m), VIS ($\lambda=0.55-1.02\mu$m) and NIR ($\lambda=1.0-2.5\mu$m). Observations were performed on 19 January and 18-19 February 2018 with clear conditions and $0.6-0.8''$ seeing. GD71 was observed as telluric standard star. We used a 1.0$''$ slit in the UVB arm and 0.9$''$ slits in the VIS and NIR arms. 

We first acquired on a $I=13.5$ magnitude star and then applied a blind offset of 95.86$''$ to the position of COLA1 with a position angle of 0 degrees (see Fig. $\ref{fig:thumbnails}$). We nodded between two positions along the slit (A and B, offset by 3$''$) using the \texttt{AutoNodOnSlit} procedure. Integration times were 700 and 780s for UVB and VIS arms respectively. Four shorter jitters with separations of 1$''$ along the direction of the slit were used for NIR exposures of 210s. This observing sequence was repeated ten times (split over two observing blocks with ABBA sequence and one with an AB sequence), resulting in total exposure times of 7.0, 7.8 and 8.4ks for the three arms respectively.

Data are reduced using standard procedures (bias and dark subtraction, flat-fielding, sky subtraction, wavelength calibration and flux calibration) incorporated in the X-SHOOTER pipeline \citep{Modigliani2010}. Individual exposures per observing block are combined using the pipeline. We then combine exposures over multiple observing blocks by computing the noise-weighted average as described in detail in \cite{Sobral2018b}. Before extracting 1D spectra, we first smooth the spectrum with 2D gaussian kernels that corresponds to half the resolution ($\sigma_x, \sigma_y$)$_{\rm UVB, VIS}$ = (0.4 {\AA},  0.32$''$) and ($\sigma_x, \sigma_y$)$_{\rm NIR}$ = (0.6 {\AA},  0.21$''$), and bin it in the wavelength direction by 3 pixels (0.6 {\AA}, half the resolution). The spatial extraction window in the VIS arm is 2.2$''$ which we find to optimise the S/N. Slit losses of the line at 922 nm are estimated using the NB921 image that is convolved to the PSF-FWHM of our spectroscopic observations. We measure the fraction of the total flux that is retrieved within the slit and extraction window. As COLA1's line-emission is compact in the NB921 data, the estimated slit losses are only 19 \%. The spatial extraction windows in the UVB and NIR arms are 1.2$''$ and we do not apply slit loss corrections. Wavelengths are converted to vacuum wavelengths. We measure the effective spectral resolution using unresolved skylines on our extracted 1D spectrum. We find $\rm R\approx4000$ at 0.5$\mu$m, $\rm R\approx 5000$ at $0.9\mu$m and $\rm R\approx3800$ at $1.6\mu$m, corresponding to 75, 60 and 80 km s$^{-1}$, respectively.

\subsection{Spectroscopic measurements}
In our X-SHOOTER spectrum, we confirm COLA1's double peaked emission line at $\lambda_{\rm obs, vac} = 9224, 9231$ {\AA} ($\rm S/N\approx24$), but we also detect faint emission-lines at $\lambda_{\rm obs, vac}=3821$ {\AA} ($\rm S/N\approx10$) and $\lambda_{\rm obs, vac}=15735$ {\AA} ($\rm S/N\approx5$). The centroid of these faint emission-lines is however shifted spatially by 1$''$ to the west and are therefore not co-located with COLA1 (Fig. $\ref{fig:thumbnails}$). We do not detect continuum emission. No other lines are detected in the $0.3-2.5\mu$m coverage above $\rm S/N>2$.

For the double peaked emission-line at 923nm, we measure a line-flux of $5.90\pm0.24\times10^{-17}$ erg s$^{-1}$ cm$^{-2}$ and a relative flux ratio between the blue and red component of $0.31\pm0.03$. The red line has a full width half maximum (FWHM) of $198\pm14$ km s$^{-1}$, while the blue line is narrower with FWHM=$150\pm18$ km s$^{-1}$. While the red line is clearly asymmetric (with a weighted skewness of $S_w=18.0\pm0.9$ {\AA}; following the definition from \citealt{Kashikawa2006}), the blue line is not ($S_w=-0.2\pm0.3$ {\AA}). Our measurements are summarised in Table $\ref{tab:measurements}$. The spectroscopically measured flux differs from the narrow-band measurement at the $1.3\sigma$ level, potentially due to systematic uncertainty in the flux calibration or the aperture correction. For the remainder of the paper we use the narrow-band flux corrected for the filter transmission.

The line at 3821 {\AA} has a line-flux of $3.0\pm0.3\times10^{-17}$ erg s$^{-1}$ cm$^{-2}$ and FWHM$=350$ km s$^{-1}$. The line that is observed at 15735 {\AA} has a line-flux of $1.5\pm0.3\times10^{-17}$ erg s$^{-1}$ cm$^{-2}$ and a width FWHM$=260$ km s$^{-1}$. The spatial offset of $\approx1''$ of these lines coincides with ID 593625.

\begin{table}
\centering
\caption{The details of the X-SHOOTER observations and spectroscopic measurements for COLA1. Flux measurements of the 923nm line are corrected for 19 \% slit-losses. Upper and lower limits are at the 2$\sigma$ level. Flux limits assume a line-width FWHM=200 km s$^{-1}$.} \label{tab:measurements}
\begin{tabular}{lc} \hline
\bf Observations & \bf COLA1 \\
Dates & 19 Jan 2018, 18-19 Feb 2018 \\
$R_{0.9 \rm \mu m}$ & 5000 \\
$R_{1.6 \rm \mu m}$ & 3800 \\
t$_{\rm exp,\, UVB}$ & 7.0ks \\
t$_{\rm exp,\, VIS}$ & 7.8ks \\
t$_{\rm exp,\, NIR}$ & 8.4ks \\
\hline
\bf 923 nm line properties & \\
Flux$_{\rm spec}$ & $5.90\pm0.24\times10^{-17}$ erg s$^{-1}$ cm$^{-2}$ \\
Flux$_{\rm NB}$ & $8.1\pm1.6\times10^{-17}$ erg s$^{-1}$ cm$^{-2}$ \\
EW$_{\rm obs}$ & $900^{+340}_{-240}$ {\AA} \\ 
Peak separation & $220\pm20$ km s$^{-1}$ \\
FWHM$_{\rm red}$ & $198\pm14$ km s$^{-1}$ \\
FWHM$_{\rm blue}$ & $150\pm18$ km s$^{-1}$ \\
Flux$_{\rm blue}$/Flux$_{\rm red}$ & $0.31\pm0.03$ \\
$S_{w,\, \rm red}$ & $18.0\pm0.9$ {\AA} \\
$S_{w,\, \rm blue}$ & $-1.5\pm0.4$ {\AA} \\
$S_{w,\, \rm full}$ & $3.2\pm0.2$ {\AA} \\ \hline
\bf Flux limits of interest & \\
H$\alpha_{z=1.475}$ & $<5.2\times10^{-18}$ erg s$^{-1}$ cm$^{-2}$ \\
{[O{\sc iii}]}$_{4959,\, z=1.475}$ & $<9.8\times10^{-18}$ erg s$^{-1}$ cm$^{-2}$ \\
{[O{\sc iii}]}$_{5007,\, z=1.475}$ & $<9.0\times10^{-18}$ erg s$^{-1}$ cm$^{-2}$ \\
C{\sc iv}$_{z=6.591}$ & $<1.6\times10^{-17}$ erg s$^{-1}$ cm$^{-2}$ \\
He{\sc ii}$_{z=6.591}$ & $<0.7\times10^{-17}$ erg s$^{-1}$ cm$^{-2}$ \\ \hline
\end{tabular}
\end{table}

\section{Is COLA1 a LAE at $z=6.59$ or an [O{\sc ii}] emitter at $z=1.47$?}\label{sec:analysis}
The tentative detection in the $B$ band (although at the $2.4\sigma$ level) and the unexpected observation of Ly$\alpha$ flux blue-wards of the red asymmetric line might suggest that COLA1 may not be a LAE at $z=6.6$, but an [O{\sc ii}] emitter at $z=1.47$ instead. Here, we assess the implications of each observed feature to test this hypothesis.

\begin{figure*}
\centering
	\includegraphics[width=14cm]{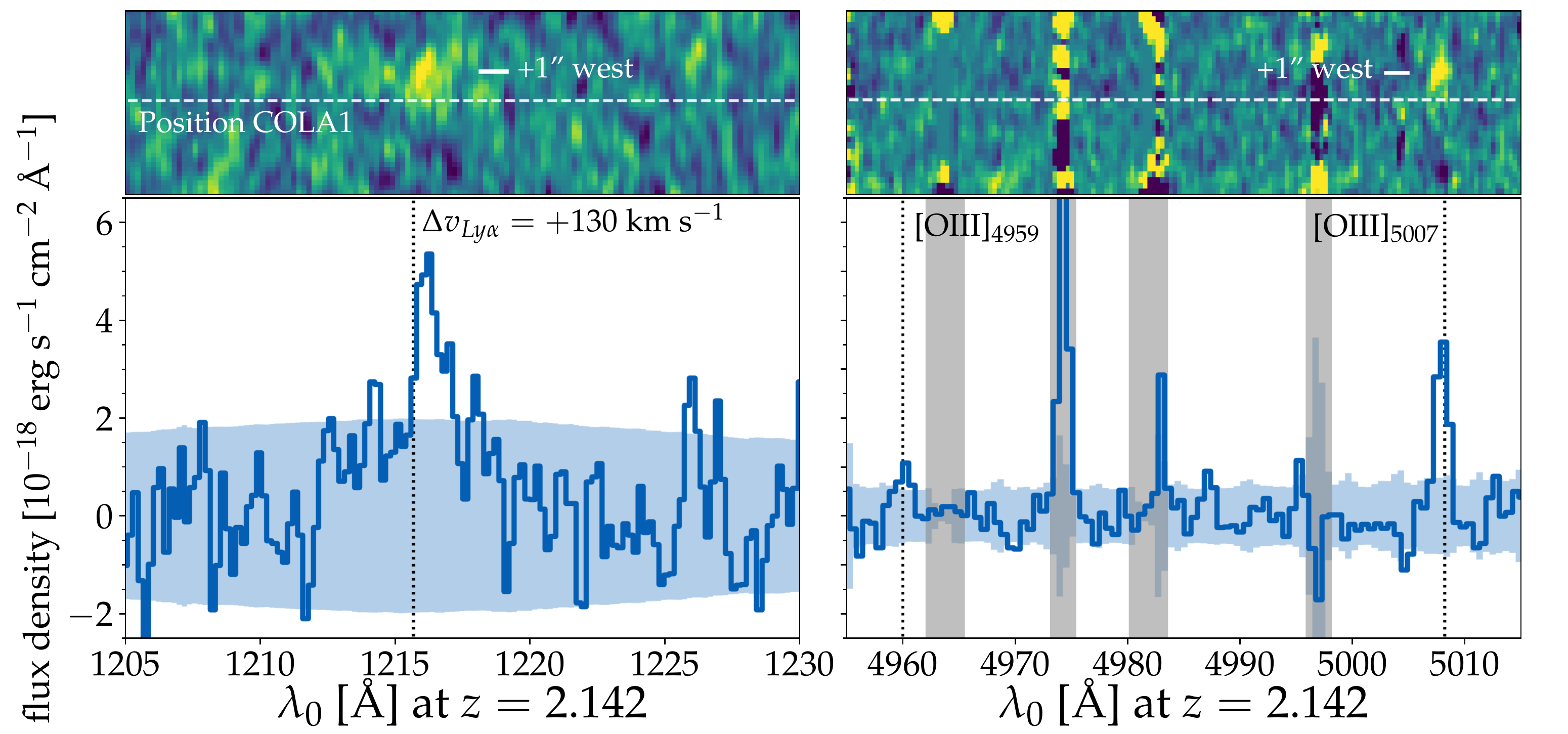} 
    \caption{Ly$\alpha$ and [O{\sc iii}]$_{5007}$ detections in the X-SHOOTER spectrum that are offset by $\approx1''$ to the west. This confirms that a foreground source (ID 593625 in \citealt{Laigle2016}) is present at $z=2.142$. Ly$\alpha$ emission is redshifted by $+130$ km s$^{-1}$ compared to the systemic redshift based on [O{\sc iii}]$_{5007}$. }
    \label{fig:foreground_spectrum}
\end{figure*}

\subsection{923nm line strength and consistency with F814W} 
If the 923nm line from COLA1 is [O{\sc ii}], it would yield $z=1.475$ and it would have a high luminosity (L$_{\rm [OII]} \approx 2\times L^{\star}$; \citealt{Khostovan2015}). The [O{\sc ii}] equivalent width would be extreme, EW$_0=360^{+140}_{-100}$ {\AA}, well above the typical EW for [O{\sc ii}] emitters (EW$_0 =50\pm20$ {\AA}) at $z=1.47$ \citep{Khostovan2016}. 
 
We test whether the F814W photometry can be explained by pure line-emission. Assuming negligible contribution from continuum emission and by spreading the line-flux homogeneously over the full transmission region of F814W results in F814W$=26.5\pm0.2$. Hence, the F814W photometry (we measure F814W$=26.6^{+0.4}_{-0.3}$) can be perfectly explained by pure line-emission and does not indicate flux blue-wards of the emission-line. Therefore, the F814W detection does not rule out Ly$\alpha$ at $z=6.6$.

\subsection{Line-detections at 382nm and 1573.5nm} 
As described above, we detect two emission-lines that are offset by $\approx1''$ to the west of COLA1. This corresponds roughly to the foreground source ID 593625 as can be seen in the $B$ band thumbnail image (Fig. $\ref{fig:thumbnails}$), which shows that outskirts of this galaxy fall in the X-SHOOTER slit. We identify the two lines as Ly$\alpha$ and [O{\sc iii}]$_{5007}$ at $z=2.142$, as illustrated in Fig. $\ref{fig:foreground_spectrum}$, where we have optimised the centroid of spatial extraction aperture for these lines. Ly$\alpha$ is redshifted by $130$ km s$^{-1}$ with respect to the systemic redshift, similar to other LAEs at $z\sim2$ \citep{Trainor2015,Sobral2018b}. Without correcting for slit losses, we measure a Ly$\alpha$ luminosity of $1.0\pm0.1\times10^{42}$ erg s$^{-1}$, well below the typical Ly$\alpha$ luminosity at $z\approx2.2$ ($\approx 0.4\times \rm L^{\star}$; \citealt{Sobral2016}). The $B$ band magnitude corresponding to pure line-emission at this flux-level is $B=28.2\pm0.1$. Ly$\alpha$ emission extends to close to the peak position of COLA1 and could thus contribute significantly to the faint $B$ band detection ($B=28.4^{+0.6}_{-0.4}$) at the COLA1 position. The [O{\sc iii}]$_{5007}$ luminosity is $0.5\pm0.1\times10^{42}$ erg s$^{-1}$. We note that the centroid of the (low S/N) flux in the $B$- near COLA1 is shifted slightly towards a faint $H$ band detection that could be explained by this [O{\sc iii}] flux. Therefore, several detections in the images (in particular $B$ and $H$) can be attributed to LAE 593625 at $z=2.142$.

\begin{figure}
	\includegraphics[width=9cm]{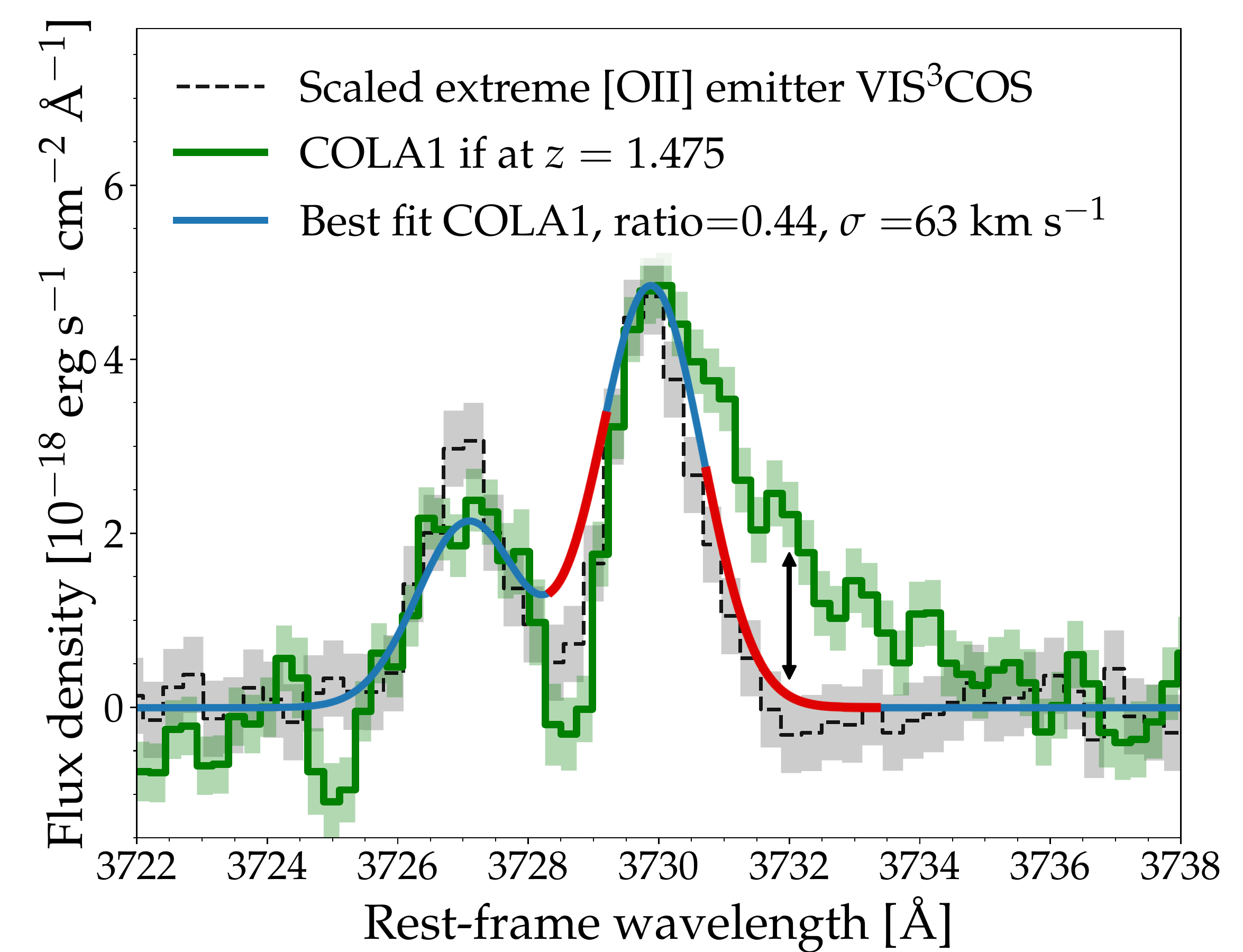} 
    \caption{X-SHOOTER spectrum of COLA1, shifted to rest-frame wavelengths assuming it is an [O{\sc ii}] emitter at $z=1.475$ (green). Dashed black lines show the median [O{\sc ii}] profile of the [O{\sc ii}] emitters in the VIS$^3$COS survey (\citealt{PaulinoAfonso2018VIS}) that are selected on having extreme blue/red ratios and asymmetric red lines. The blue line shows our fitted [O{\sc ii}] doublet to the spectrum of COLA1, and the red line indicates where the fit is $>3\sigma$ away from the data. The asymmetry in the red line (not present in the blue line) and the absence of flux in between the lines are at odds with COLA1 being an [O{\sc ii}] emitter.   } 
        \label{fig:visspec_ifO2}
\end{figure} 

\subsection{923nm line-profile} 
We show a detailed zoom-in figure of the line-profile of the 923nm line in Fig. $\ref{fig:visspec_ifO2}$. In this figure, we shift the spectrum to the rest-frame assuming COLA1 is at $z=1.475$. We compare it to a median [O{\sc ii}] spectrum of galaxies with asymmetric red lines and high red/blue ratios from the VIS$^3$COS survey \citep{PaulinoAfonso2018VIS} and also show the best fitted [O{\sc ii}] doublet. While the peak separation ($220\pm20$ km s$^{-1}$) is fully consistent with the peak separation of the [O{\sc ii}] doublet, the asymmetry of the red line can not be fitted as an [O{\sc ii}] doublet. In particular, the blue line would have been expected to be asymmetric as well, similar to the red line. Moreover, the absence of flux between the two lines can also not easily be explained in the case of an [O{\sc ii}] doublet, unless lines are very narrow. 

The skewness of COLA1's red line is high ($S_{w,\, \rm red} = 18.0\pm0.9$ {\AA}; see Table $\ref{tab:measurements}$), much higher than the typical maximum skewness of a low-redshift galaxy ($S_w=3$ {\AA}; \citealt{Kashikawa2006}), which is similar to the skewness of the full doublet. 
Finally, the line-ratio between [O{\sc ii}]$_{3726}$ and [O{\sc ii}]$_{3730}$ would be significantly lower than the line-ratio in our extreme [O{\sc ii}] emitter reference sample (with a blue-to-red fraction of $>0.65$). In fact, the line-ratio of the blue and red lines of our best-fit (0.44) is significantly lower than the theoretical minimum line-ratio for electron densities as low as 1 cm$^{-3}$ ($\approx0.65$ for an electron temperature of 10$^4$ K; \citealt{Sanders2016}, but also for temperatures between $10^{3-5}$ K) and thus unphysical. 
Therefore, the line-profile strongly disagrees with COLA1 being an [O{\sc ii}] emitter at $z=1.475$ even though the peak separation is in perfect agreement.

\subsection{No line in the NIR associated with $z=1.475$} 
\begin{figure}
	\includegraphics[width=9.3cm]{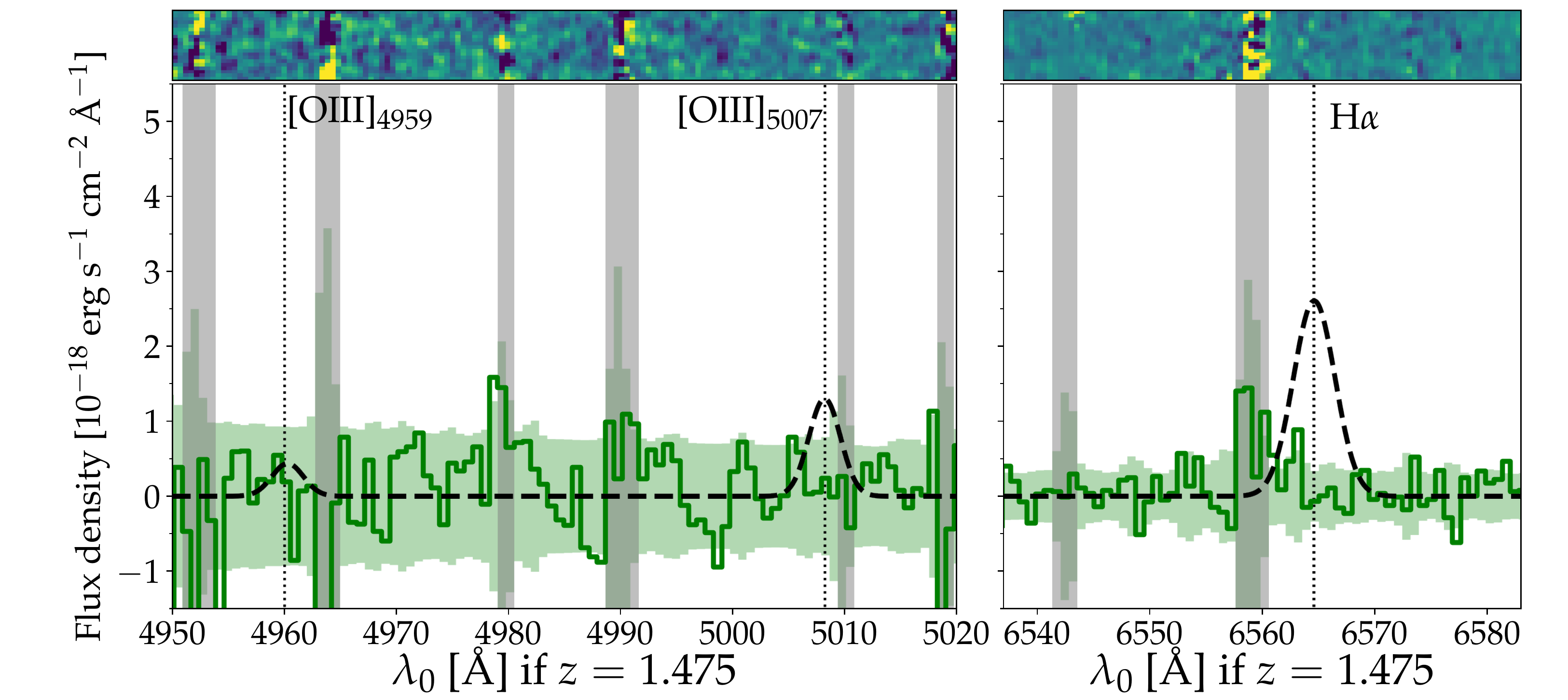} 
    \caption{The wavelengths where [O{\sc iii}] and H$\alpha$ would be observed if COLA1 is at $z=1.475$. The dashed lines indicate the minimum expected [O{\sc iii}] and H$\alpha$ fluxes based on known extreme line-ratios ([O{\sc iii}]/[O{\sc ii}]=0.1 and [O{\sc ii}]/H$\alpha$ = 0.2, shown with FWHM=200 km s$^{-1}$). These line-ratios are ruled out at $\approx1\sigma$ and 4.7$\sigma$, respectively.}
    \label{fig:nir_z1p5}
\end{figure}

As we show in Fig. $\ref{fig:nir_z1p5}$, we do not detect H$\alpha$ or [O{\sc iii}]$_{5007}$ if COLA1 would be at $z=1.475$. We illustrate how these lines would look in our data for the lowest empirical H$\alpha$/[O{\sc ii}]$=0.2$ and [O{\sc iii}]/[O{\sc ii}]$=0.1$ line-ratios \citep[e.g.][]{Hayashi2013}. These extreme line-ratios would imply both high metallicities and high ionisation states (properties that are typically anti-correlated; \citealt{NakajimaOuchi2013}). If we conservatively assume the spectroscopic flux of the 923nm line and  a FWHM of 200 km s$^{-1}$ (similar to the red line at 923nm), we measure a 1$\sigma$ limit on the H$\alpha$ flux of $2.6\times10^{-18}$ erg s$^{-1}$ cm$^{-2}$ and a lower limit of $9.4\times10^{18}$ erg s$^{-1}$ cm$^{-2}$ on the combined [O{\sc iii}]$_{4959, 5007}$ line. While the lowest [O{\sc iii}]/[O{\sc ii}] ratio is within the 1$\sigma$ noise level, these data rule out H$\alpha$/[O{\sc ii}]$=0.2$ at $4.7\sigma$, additional clear evidence against the interpretation that COLA1 is an [O{\sc ii}] line at $z=1.475$.

\subsection{Concluding remarks on the redshift of COLA1} 
As the line-profiles of the red and blue lines differ, the double peaked emission line around 923nm can not be fitted by an [O{\sc ii}] doublet, even though the peak separation is similar\footnote{As we show in Section $\ref{sec:COLA1_properties}$, the Ly$\alpha$ line profile is well-fitted by a Ly$\alpha$ shell model and shows properties similar to normal LAEs at $z<6$.}. Photometric indications for COLA1 being at low-redshift (in particular $B$ and F814W detections) are explained due to line-emission from COLA1 itself (F814W) and a foreground LAE at close separation (ID 593625 at $z=2.142$). The relatively blue {\it Spitzer}/IRAC colours, combined with a optical to near-infrared break of $BVRI-Y>3$ are also not indicative of a red dusty or old interloper at a lower redshift. Finally, if COLA1 would have been an [O{\sc ii}] emitter at $z=1.475$ the flux ratio of the lines in the [O{\sc ii}] doublet and the limits on [O{\sc ii}]/H$\alpha$ indicate unphysical conditions. 
Combining all the observations from above, we conclude that COLA1 is best explained as a double-peaked LAE at $z_{\rm Ly\alpha,red}=6.593$, as initially proposed by \cite{Hu2016} and corroborated by \cite{Songaila2018}.

\section{Properties of COLA1 -- a unique LAE at $z=6.6$} \label{sec:COLA1_properties}
Now we have established that COLA1 is a real LAE at $z=6.593$, we can have a better look at its properties based on the X-SHOOTER spectrum and available imaging data. The properties are summarised in Table $\ref{tab:lyashell}$.

\subsection{Ly$\alpha$ luminosity and spectral energy distribution} \label{sec:SED}
The transmission-corrected narrow-band magnitude implies a Ly$\alpha$ luminosity of L$_{\rm Ly\alpha}=4.1\times10^{43}$ erg s$^{-1}$, which is among the most luminous LAEs know at $z\approx5-7$ (see a compilation in \citealt{Matthee2017spec}). The Ly$\alpha$ EW is high (EW$_0=120^{+50}_{-40}$ {\AA}, based on the continuum estimated from the $Y$ band flux), but this is a rather common property of LAEs at high-redshift \citep[e.g.][]{Hashimoto2017,Sobral2018} and we note that it is poorly constrained due to uncertainties on the continuum magnitude. The Ly$\alpha$ luminosity implies a comoving number density $\approx1\times10^{-5}$ Mpc$^{-3}$ \citep{Matthee2015}.

\begin{figure}
	\includegraphics[width=9.3cm]{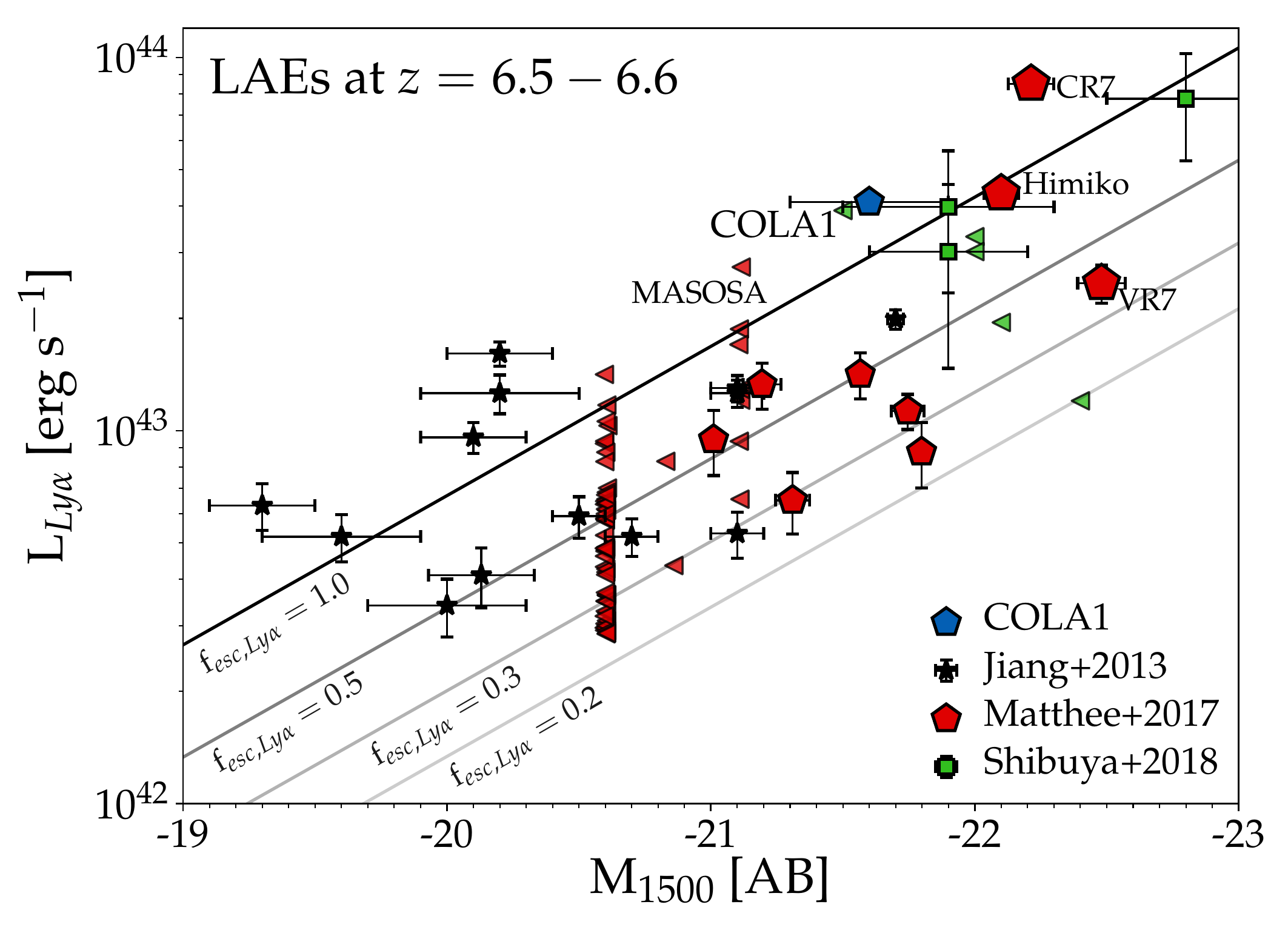} 
    \caption{UV luminosity versus Ly$\alpha$ luminosity of known LAEs at $z=6.5-6.6$ identified through narrow-band surveys (\citealt{Jiang2013_MUV,Matthee2017spec,Shibuya2017}). Left-ward pointing triangles indicate 2$\sigma$ upper limits. Black lines show lines of constant escape fraction (f$_{\rm esc, Ly\alpha}$; assuming a Salpeter IMF and a constant SFR for 100 Myr; e.g. \citealt{Kennicutt1998}). COLA1 has a high Ly$\alpha$ escape fraction, similar to other LAEs known at $z=6.6$. }
    \label{fig:cola1_UV}
\end{figure}

COLA1 is weakly detected in the $Y$ and $J$ bands ($\rm S/N \approx 3$ and $2.5$, respectively),
see Fig. $\ref{fig:thumbnails}$. An off-centred $H$ band detection is likely contaminated by strong [O{\sc iii}]$_{4959,5007}$ emission from the foreground galaxy at $z=2.14$, and we can only provide a 2$\sigma$ limit on the magnitude of $H>24.8$. COLA1 is undetected in the $K_s$ band ($K_s>24.4$). Due to the large uncertainties in the photometry no meaningful constraints can be obtained on the UV slope, and we assume a flat UV slope $\beta=-2$ in the rest of the paper (this is similar to other LAEs at $z=6.6$; \citealt{Ono2010,Matthee2017spec}).

The $Y$ band magnitude implies an absolute magnitude $M_{1500}=-21.6\pm0.3$, slightly above M$_{1500}^{\star}$ at $z\approx7$ \citep{Bouwens2014}, but fainter than other luminous LAEs at $z=6.6$ \citep[e.g. CR7 and VR7;][]{Matthee2017spec}, see Fig. $\ref{fig:cola1_UV}$. Following \cite{Shibuya2015}, such a UV luminosity indicates M$_{\rm star}\approx10^{10}$ M$_{\odot}$ at $z\sim7$, although this could be significantly underestimated in case the source is strongly dust-obscured (which however is at odds with its strong Ly$\alpha$ emission; e.g. \citealt{Matthee2016}). The narrowness of the Ly$\alpha$ line indicates that COLA1 is likely a star-forming galaxy, as AGN typically have broader Ly$\alpha$ lines \citep[e.g.][]{Matsuoka2016}. COLA1 is unlikely to be significantly magnified: no massive foreground structures are known and no distortions are seen on the images. 

COLA1's colours in the shortest {\it Spitzer}/IRAC channels are blue, $[3.6]-[4.5]=-0.2\pm0.3$. This is likely a consequence of H$\beta$ and/or [O{\sc iii}] slightly boosting the [3.6] flux more than H$\alpha$ is boosting the [4.5] flux (which is in a low transmission wavelength of the [4.5] band). However, COLA1 is not nearly as blue as other confirmed galaxies at $z\sim6.6$. For example, CR7 and Himiko have colours $[3.6]-[4.5]=-1.3\pm0.3$ and $[3.6]-[4.5]=-0.7\pm0.4$, respectively \citep{Harikane2018}. This could indicate that COLA1's [O{\sc iii}] line is relatively weaker due to a lower metallicity. When we compare this colour to the predicted colours at $z=6.6$ using photoionisation analysis and BPASS SED models presented in \cite{Bowler2017}, we find that the implied metallicity lies between $Z\approx10^{-4}-10^{-3}$, but always consistent with $Z<10^{-3} (<1/20\,Z_{\odot})$ within the uncertainties. Alternatively, the IRAC colour could also indicate that an older stellar population is present in COLA1, or that the strength of nebular lines is lower due to a high LyC escape fraction \citep[e.g.][]{Zackrisson2017}.

\begin{figure}
	\includegraphics[width=9cm]{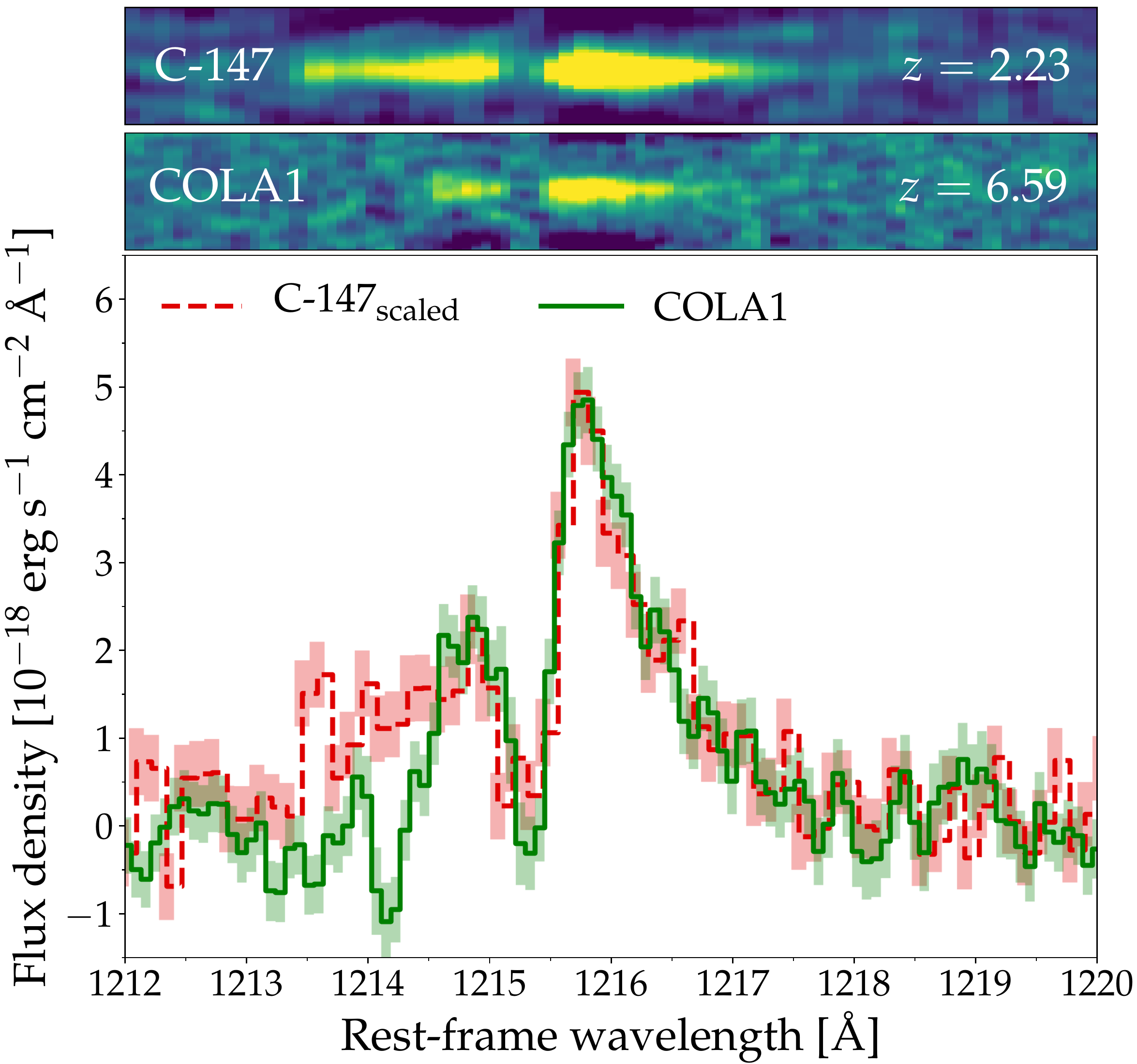} 
    \caption{X-SHOOTER spectrum of COLA1, shifted to rest-frame Ly$\alpha$ at $z=6.593$ (green), compared to spectra of C-147 (red dashed line, a double peaked LAE at $z=2.23$ confirmed through H$\alpha$ emission, scaled to the peak flux in the red line; \citealt{Sobral2018b}). The profile of the red line in COLA1 resembles the Ly$\alpha$ profile of C-147 remarkably well. The blue component of COLA1 is cut at $\Delta v\approx -250$ km s$^{-1}$, and thus narrower than C-147, potentially due to a higher optical depth at large distance from COLA1.}
    \label{fig:visspec}
\end{figure}

\subsection{Ly$\alpha$ size and star formation rate density}
As discussed in \S $\ref{sec:analysis}$, the detection of COLA1 in the F814W band can be explained by pure Ly$\alpha$ emission. As the {\it HST}/ACS imaging has a small PSF, we can therefore use these data to constrain the size of the Ly$\alpha$ emitting region in COLA1 accurately \citep[e.g.][]{PaulinoAfonso2018}. We use a non-parametric method to measure the median half-light radius (r$_{50, \rm obs}$) for 5000 random realisations of the image, where each pixel count is drawn from a Poissonian distribution. Then, we correct for PSF-broadening using $r_{50} = \sqrt{r_{50, \rm obs}^2 - r_{\rm PSF}^2}$, where $r_{\rm PSF}=0.38$ kpc. This results in $r_{50} = 0.33^{+0.07}_{-0.04}$ kpc. 
This size is smaller than typical UV-selected galaxies with M$_{1500}\approx-21$, similar to galaxies with M$_{1500}\approx-18.4$ in the UV luminosity - size relation from \cite{Shibuya2015} and similar to the largest star cluster complexes \citep{Bouwens2017,Vanzella2017}. As the UV sizes of galaxies are typically much smaller than the Ly$\alpha$ sizes \citep[e.g.][]{Wisotzki2015}, it could be possible that the galaxy is even more compact in the UV. On the other hand, it is likely that more diffuse Ly$\alpha$ emission is resolved out in the broad, high resolution F814W image (similar to e.g. in CR7; more luminous, extended Ly$\alpha$ emission that is undetected in similar F814W data), and that the core Ly$\alpha$ profile resembles the UV profile. Ly$\alpha$ emission is expected to be compact in galaxies with a low H{\sc i} column density and high f$_{\rm esc, LyC}$ \citep{MasRibas2017} -- conditions that are likely present in COLA1 as discussed next.

We can estimate the SFR of COLA1 from the UV continuum or from the Ly$\alpha$ luminosity. Assuming a UV slope $\beta=-2.0$ and a \cite{Meurer1999} attenuation law, we find a dust-corrected SFR$_{\rm UV}=40^{+12}_{-10}$ M$_{\odot}$ yr$^{-1}$ ($27^{+8}_{-7}$ M$_{\odot}$ yr$^{-1}$ without correcting for dust). Alternatively, the Ly$\alpha$ luminosity indicates SFR$_{\rm Ly\alpha} = 75^{+35}_{-25}$ M$_{\odot}$ yr$^{-1}$ (assuming a 60 \% Ly$\alpha$ escape fraction and 15 \% LyC escape fraction; following \citealt{SobralMatthee2018}). 

If we assume that the UV size is similar to the Ly$\alpha$ size, these measurements imply a minimum average SFR surface density $\Sigma_{\rm SFR, UV, no dust} = 95^{+50}_{-35}$ M$_{\odot}$ yr$^{-1}$ kpc$^{-2}$. This number would be higher if the UV size is more compact or if significant dust attenuation is present. This $\Sigma_{\rm SFR}$ is significantly higher than other galaxies known at $z>6$ \citep[e.g.][]{Carniani2018}. The SFR surface density is well above the threshold required to drive galactic outflows \citep[e.g.][]{Heckman2001}, hence such outflows may well be present.
\begin{table}
\centering
\caption{Derived properties of COLA1 as described in \S $\ref{sec:COLA1_properties}$. Limits are at the 2$\sigma$ level.} \label{tab:lyashell}
\begin{tabular}{lc} \hline
\bf Property & \bf Best estimate\\ \hline
Spectral analysis & \\
L$_{\rm Ly\alpha}$ & $4.1\pm0.2\times10^{43}$ erg s$^{-1}$ \\
EW$_{0, \rm Ly\alpha}$ & $120^{+50}_{-40}$ {\AA} \\
M$_{1500}$ & $-21.6\pm0.3$ \\
M$_{\rm star}$ & $\approx 10^{10}$ M$_{\odot}$ \\
$r_{50}$ & $0.33^{+0.07}_{-0.04}$ kpc \\
$f_{\rm esc, LyC}$ & $\approx15-30$ \% \\
SFR$_{\rm Ly\alpha}$ & $75^{+35}_{-25}$ M$_{\odot}$ yr$^{-1}$ \\
SFR$_{\rm UV, no dust}$ & $27^{+8}_{-7}$ M$_{\odot}$ yr$^{-1}$\\
$\Sigma_{\rm SFR, UV, no dust}$ & $95^{+50}_{-35}$ M$_{\odot}$ yr$^{-1}$ kpc$^{-2}$\\ 
Gas-phase metallicity $Z$ & $\lesssim10^{-3} (1/20\,Z_{\odot})$\\
\hline
Shell model fitting& \\
v$_{\rm exp}$ & $78^{+5}_{-6}$ km s$^{-1}$\\
log$_{10}$(N$_{\rm HI}$/cm$^{-2})$&$17.0^{+0.3}_{-0.3}$\\
log$_{10}$(T/K) & $4.2^{+0.1}_{-0.2}$\\
$\sigma_{\rm intrinsic}$ & $159^{+4}_{-4}$ km s$^{-1}$\\
$\tau_d$ & $4.2^{+0.5}_{-0.8}$\\
$z_{\rm sys}$ & $6.5930^{+0.0001}_{-0.0002}$\\
\hline
\end{tabular}
\end{table}

\subsection{Ly$\alpha$ line properties and modelling}
In Fig. $\ref{fig:visspec}$ we show COLA1's rest-frame spectrum assuming $z=6.593$ (the redshift of the red peak) and compare it to the luminous LAE C-147 (a Ly$\alpha$-selected star-forming galaxy at $z=2.23$ for which a blue peak is also detected in the Ly$\alpha$ spectrum and the systemic redshift is confirmed through H$\alpha$ and [O{\sc iii}] emission; \citealt{Sobral2016,Sobral2018b}). The red part of the Ly$\alpha$ line of C-147 resembles the red asymmetric line in COLA1 very well. C-147 has a broader blue line, but the peak separation and the fact that there is no significant flux at line-centre is similar to COLA1. 

If we assume that the transmitted Ly$\alpha$ flux is zero at line-centre (in between the two peaks; e.g. \citealt{Yang2017}), we measure a systemic redshift $z=6.591$ for COLA1. The peak separation of the blue and red Ly$\alpha$ lines is $220\pm20$ km s$^{-1}$. In LyC leaking galaxies in the local Universe, the peak separation is anti-correlated with the amount of leakage of LyC photons \citep{Verhamme2015,Verhamme2017}. If this correlation would hold up to $z=6.6$,  we would infer f$_{\rm esc, LyC} \approx 15$ \% for COLA1, but potentially up to f$_{\rm esc, LyC} \approx 30$ \% if we compare to the most recent results \citep{Izotov2018}. A comparison of the blue-to-red flux-ratio of local LyC leakers also indicates f$_{\rm esc, LyC} \approx 15$ \% (although we note that the IGM could absorb part of the blue line-flux, see \S $\ref{sec:discussion}$). For the rest of the paper we thus conservatively assume f$_{\rm esc, LyC} \approx 15$ \%. Therefore, the likely non-zero LyC escape fraction of COLA1 presents the best direct evidence to date that star-forming galaxies contributed to the reionisation of the Universe at $z\sim7$. In fact, as we discuss in \S $\ref{sec:discussion}$, such a contribution is required to ionise a large enough region to be able to detect the blue peak at its redshift. The prospects for directly detecting LyC photons at $z=6.6$ are very low because of the typical opacity in the IGM is high at $z>6$ \citep{Inoue2014}. Moreover, while foreground contamination is a major issue \citep{Siana2015}, measuring the Ly$\alpha$ peak separation may be a promising alternative.

\subsubsection{Ly$\alpha$ spectral modelling}
In order to get a more quantitative view of the ISM properties implied by the observed Ly$\alpha$ line, we fit the Ly$\alpha$ profile of COLA1 using a five-parameter shell model as in \cite{Dijkstra2016}. This model \citep[e.g.][]{Ahn2001,Verhamme2006} consists of a shell of neutral gas and dust around a central ionising source and parameters include the H{\sc i} column density, expansion velocity, temperature and dust optical depth of the shell and the intrinsic width of the Ly$\alpha$ line \citep{Gronke2015}. This fit does not include transmission through the IGM. 

The best fitted parameters are listed in Table $\ref{tab:lyashell}$. Compared to $z\approx3-5$ LAEs \citep{Gronke2017}, these parameters indicate a relatively low H{\sc i} column density N$_{\rm HI}=10^{17.0\pm0.3}$ cm$^{-2}$ and a high dust optical depth $\tau_d = 4.2^{+0.5}_{-0.8}$, while the other parameters are quite common for LAEs. The low H{\sc i} column density is inferred from the low peak separation and suggests the possible escape of LyC photons. The high dust optical depth could alternatively be interpreted due to IGM opacity that could lower and narrow the blue peak. The H{\sc i} column density and expansion velocity are also significantly lower than the inferred column density around CR7 \citep[e.g.][]{Dijkstra2016}. We note that the shell-model prediction of the systemic velocity of CR7 presented in \cite{Dijkstra2016} agrees perfectly with recent [C{\sc ii}] measurements in \cite{Matthee2017ALMA}, suggesting that shell-model fitting is a good tool to recover the systemic redshift of a LAE at high redshift. The column density is also significantly lower than the column density inferred from absorption line measurements in local LyC leakers \citep{Gazagnes2018}, which lead these authors to conclude that the H{\sc i} covering fraction in these galaxies is non-uniform. A relatively high column density in these local LyC leakers is consistent with their larger Ly$\alpha$ peak separation compared to COLA1, and their typical escape fraction of $\approx 5$ \%.

As detailed in \cite{Verhamme2015}, the Ly$\alpha$ line profile can be used as a tracer of $f_{\rm esc, LyC}$ as both are sensitive to H{\sc i} column density as N$_{\rm HI}=-\ln(f_{\rm esc, LyC})/\sigma_{0}$ where $\sigma_0=6.3\times10^{-18}$ cm$^2$ is the ionisation cross section. According to this equation, the column density derived from the shell-model fit (see Table $\ref{tab:lyashell}$) implies $f_{\rm esc, LyC} \approx50$ \%. This is consistent with the result one would infer by using the correlation between Ly$\alpha$ EW$_0$ and f$_{\rm esc, LyC}$ at $z\sim3$ \citep{Steidel2018}, although we note this correlation has not been tested beyond EW$_{0} \gtrsim 50$ {\AA} and likely breaks down due to a reduced Ly$\alpha$ strength in the high f$_{\rm esc, LyC}$ regime. The Ly$\alpha$ EW also points towards a 30 \% escape fraction as implied by comparison with the most recent measurements in local LyC leakers \citep{Izotov2018}. Such high escape fractions would also affect the strength of nebular emission lines such as H$\alpha$ and H$\beta$ and could therefore be tested with future spectroscopic observations with the {\it James Webb Space Telescope} \citep[e.g.][]{Jensen2016,Zackrisson2017}. Indeed, the fact that COLA1 shows a $[3.6]-[4.5]$ colour that is much closer to zero than other luminous LAEs at the same redshift could potentially indicate a reduced strength of nebular emission lines, see \S $\ref{sec:SED}$.

\begin{figure}
	\includegraphics[width=9.3cm]{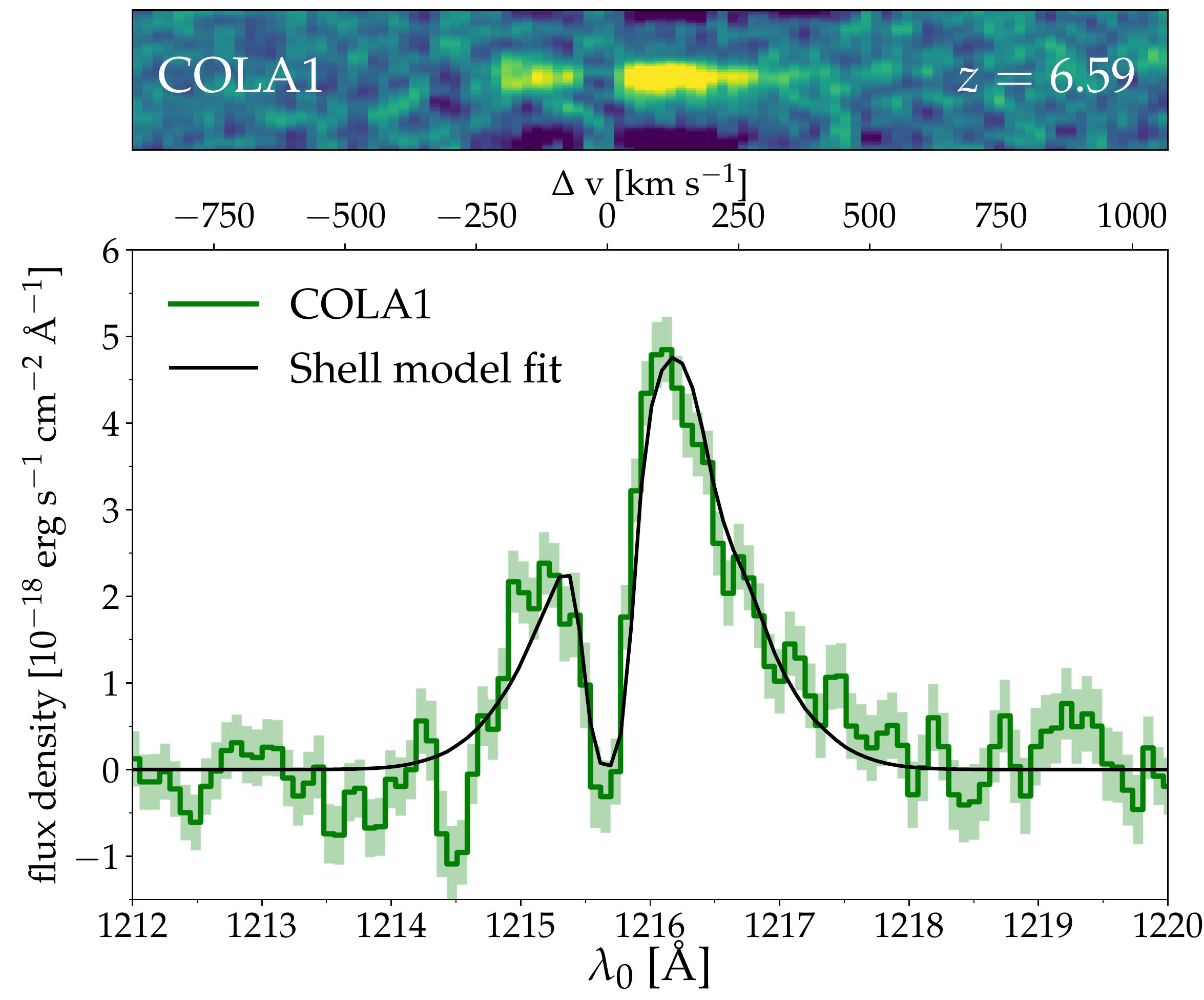} 
    \caption{The observed Ly$\alpha$ profile of COLA1 (green) is well modelled by an expanding shell of neutral gas (black). The shell-model parameters indicate a low neutral hydrogen column density N$_{\rm HI}=10^{17.0\pm0.3}$ cm$^{-2}$ and low expansion velocity v$_{\rm exp}=78^{+5}_{-6}$ km s$^{-1}$. We note that we have assumed here that the systemic redshift lies at line-centre (i.e. between the red and blue peak, where the flux is consistent with zero), or z$_{\rm Ly\alpha}=6.591$.}
    \label{fig:cola1_lya}
\end{figure}

\begin{figure}
	\includegraphics[width=9.3cm]{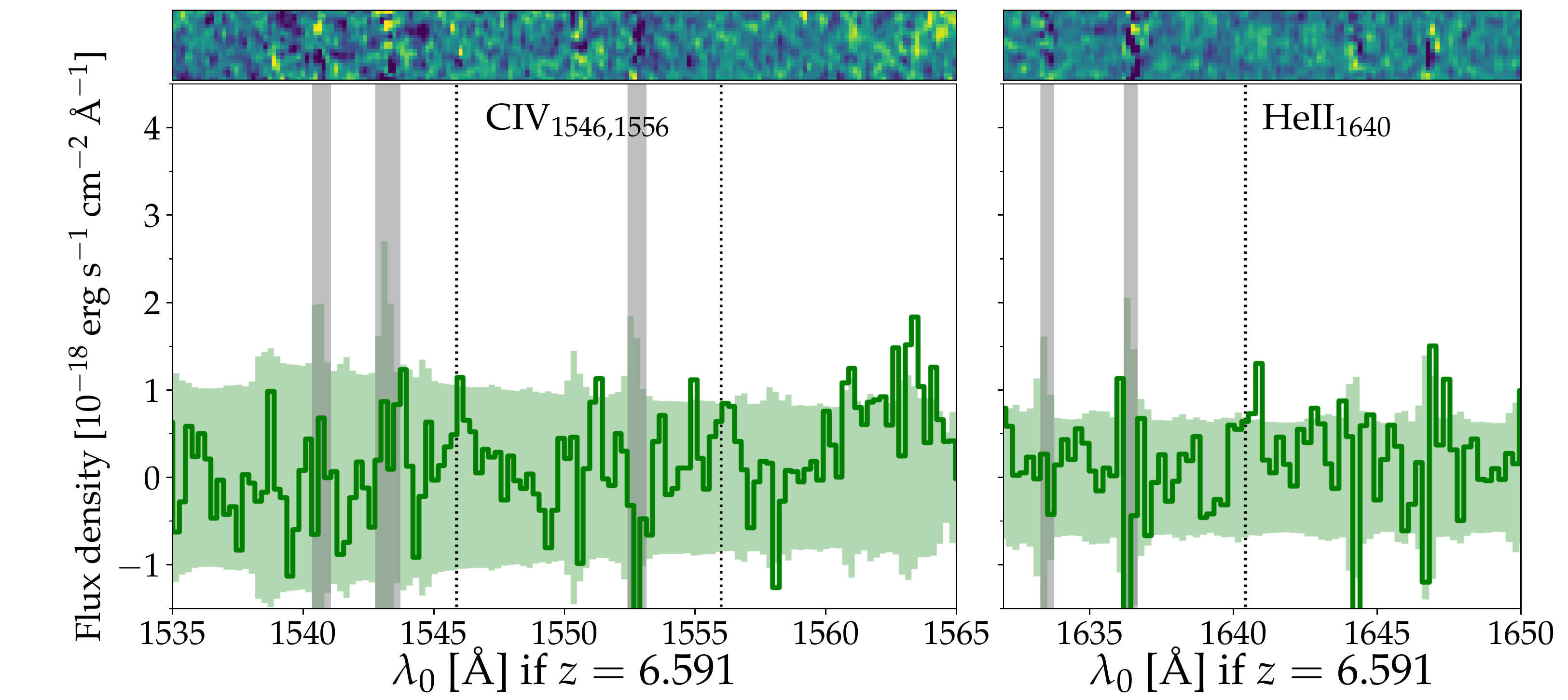} 
    \caption{The wavelength coverage of the rest-UV C{\sc iv} and He{\sc ii} lines, redshifted to $z=6.591$. We note that the background in a larger region around $\lambda_{0, z=6.591} = 1560-1565$ {\AA} is boosted due to low atmospheric transmission and faint skylines. C{\sc iv} and He{\sc ii} lines are not significantly detected in COLA1, implying EW$_{0,\rm CIV}<25$ {\AA} and EW$_{0,\rm HeII}<12$ {\AA}.}
    \label{fig:nir_z6p6}
\end{figure}

\subsection{High ionisation rest-UV emission-lines}
Thanks to the wavelength coverage of the X-SHOOTER spectrum, we can inspect the spectrum for the presence of strong rest-UV emission line features \citep[e.g.][]{Stark2015_CIII}; see Fig. $\ref{fig:nir_z6p6}$. We do not detect a significant feature besides Ly$\alpha$. We measure 2$\sigma$ limiting line-fluxes of $1.6$ and $0.7\times10^{17}$ erg s$^{-1}$ cm$^{-2}$ for the C{\sc iv}$_{1546,1556}$ and He{\sc ii}$_{1640}$ emission lines at $z=6.591$, respectively (over 250 km s$^{-1}$ extraction boxes), corresponding to EW$_{0,\rm CIV}<25$ {\AA} and EW$_{0,\rm HeII}<12$ {\AA}. These non-detections are not surprising given the fact that the limits are shallower than other unsuccessful spectroscopic follow-up of LAEs at $z=6.6$ \citep[e.g.][]{Shibuya2017}, and EW limits are higher than detections in e.g. \cite{Stark2017}.

\begin{figure*}
\centering
\begin{tabular}{ccc}
\hspace{4mm}	\includegraphics[width=5.67cm]{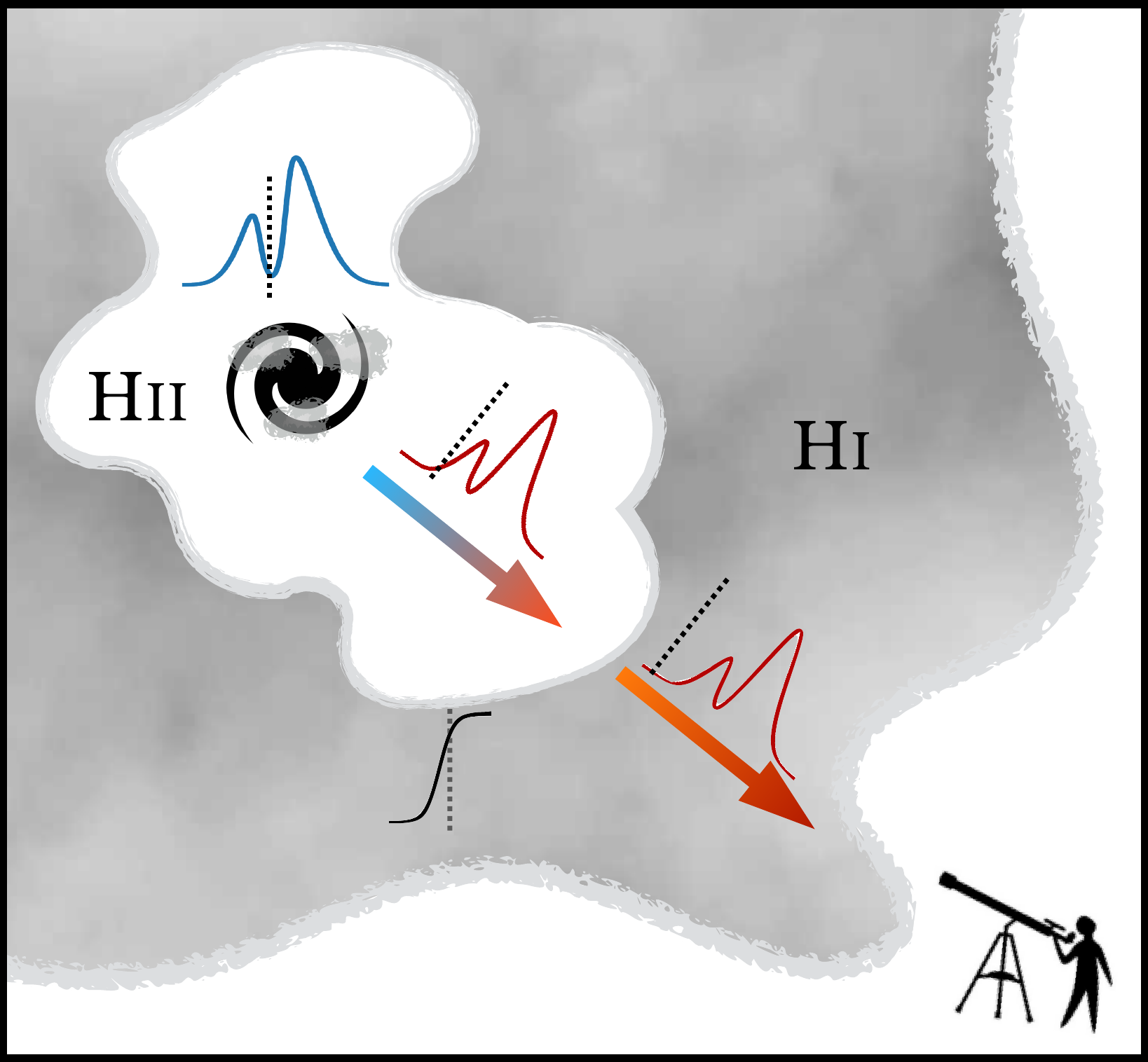}& 
\hspace{-1.9mm}	\includegraphics[width=5.67cm]{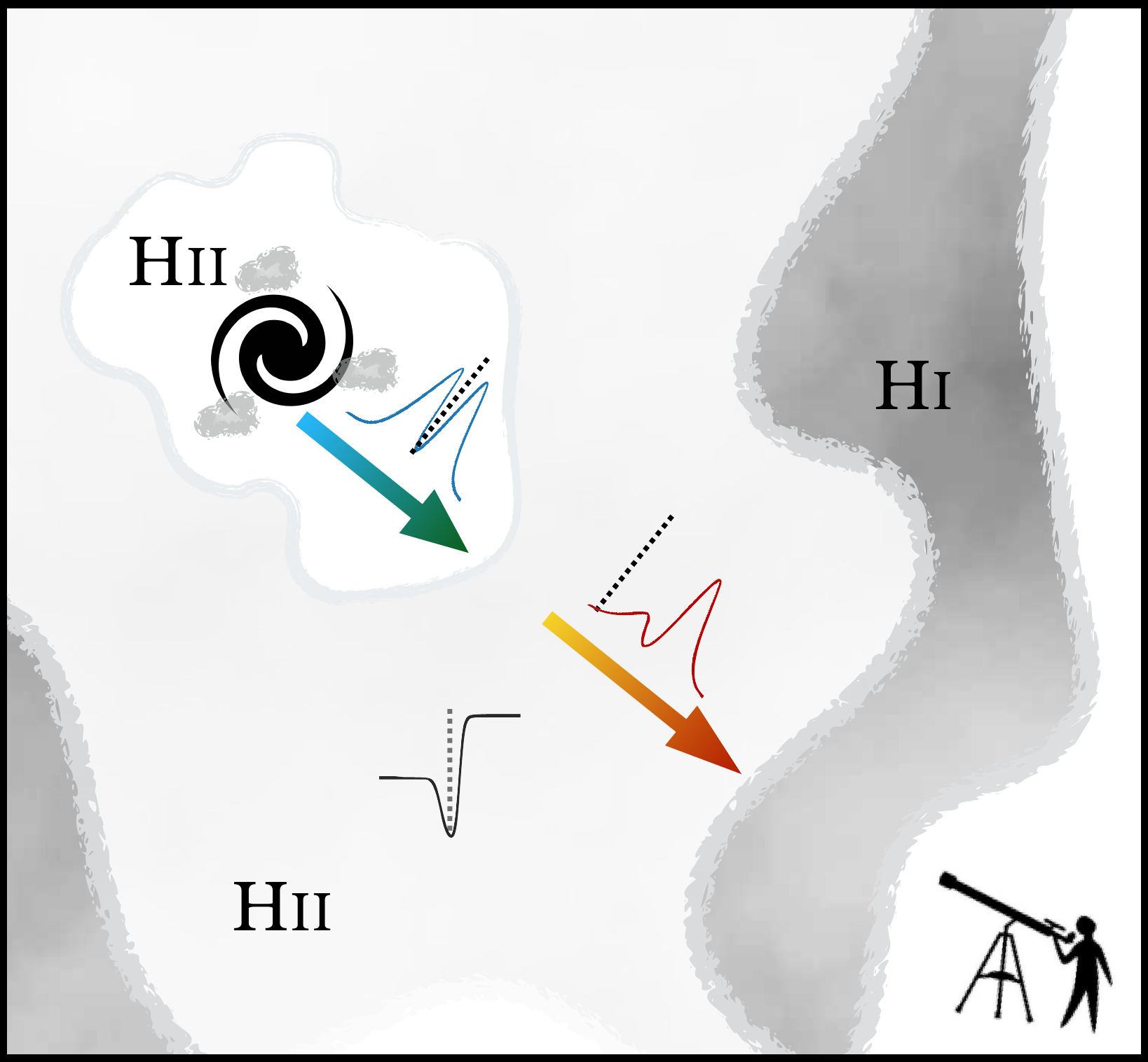} &
\hspace{-1.9mm}	\includegraphics[width=5.67cm]{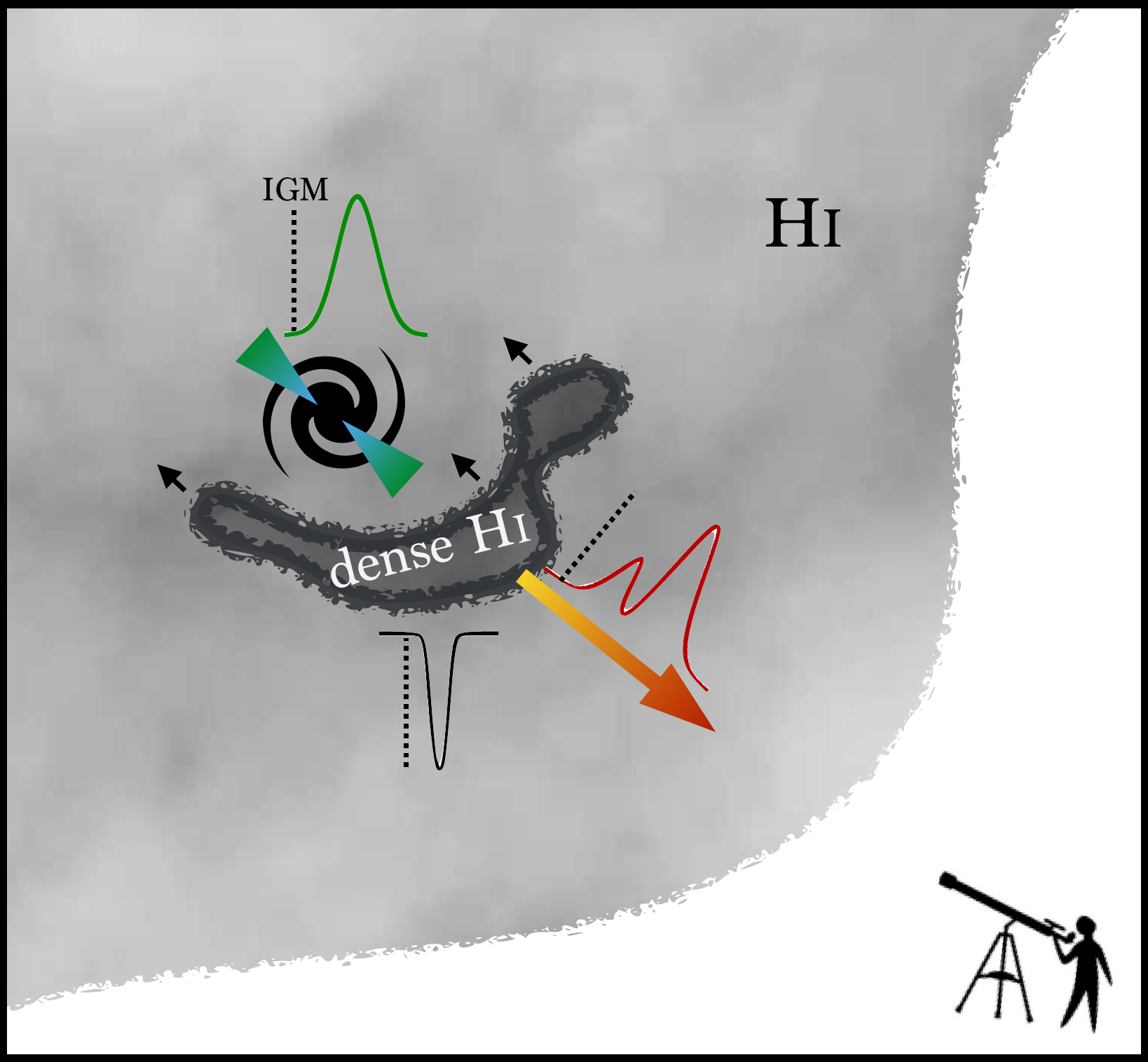} \\
\end{tabular}
	\includegraphics[width=18.6cm]{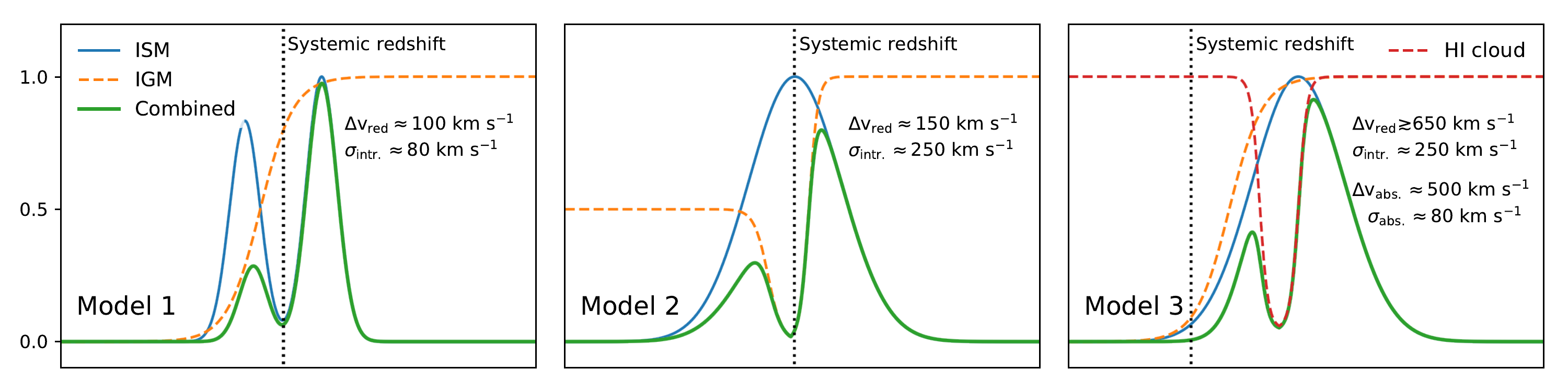} 
    \caption{Sketches of different scenarios that may explain the observed Ly$\alpha$ line-profile of COLA1. The top row shows sketches of the physical scenarios and the bottom row shows the emerging Ly$\alpha$ spectrum.  On the left, Model 1 shows a double peaked emission line emerging from scattering in the ISM that is redshifted due to the expansion of the Universe before encountering the IGM, that preferentially attenuates the blue component. In the middle, Model 2 shows that a double peak may originate due to a low transmission at line centre (due to neutral hydrogen in the CGM), while there is a relatively high transmission at further distances from the galaxy due to a large relatively ionised region (similar to the IGM transmission curve at $z\approx4$ in \citealt{Laursen2011}). On the right, Model 3 shows that a double peaked profile can also arise without IGM attenuation in case there is an H{\sc i} absorber slightly blue shifted with respect to the systemic Ly$\alpha$ velocity (which in this model is significantly redshifted with respect to the IGM).}
    \label{fig:igmillustration}
\end{figure*}

\section{Discussion: witnessing a galaxy reionising its surroundings} \label{sec:discussion}
\subsection{The unlikelihood of observing a blue Ly$\alpha$ line at $z>6$} 
Ly$\alpha$ photons resonantly scatter in the presence of neutral hydrogen. Once Ly$\alpha$ photons are absorbed, they are re-emitted in a random direction in the rest-frame of the absorbing hydrogen atom, resulting in a diffusion process in real and frequency space. Analytical models show that this process results in a double peaked Ly$\alpha$ spectrum in a static medium \citep[e.g.][]{Neufeld1990}. In the presence of outflows, Ly$\alpha$ photons see a larger optical depth towards the blue, resulting in a redshifted asymmetric spectrum. Hence, the Ly$\alpha$ profile is sensitive to the neutral hydrogen content and the velocity field of the gas in a galaxy \citep[e.g.][]{LoebRybicki1999,Santos2004,Dijkstra2007}. 

After escaping from the ISM, Ly$\alpha$ photons are also affected by H{\sc i} in the circum-galactic medium (CGM). The CGM predominantly transmits red Ly$\alpha$ photons, enhancing the asymmetry between the red and the blue peak \citep[e.g.][]{Laursen2011}. This is consistent with observations at high-redshift ($z>2$), where high H{\sc i} column densities in the CGM result in a majority of Ly$\alpha$ profiles that consist of a single, red asymmetric line \citep[e.g.][]{Erb2014}. Due to the increasing neutral fraction of the IGM, the asymmetry between the transmission of the red and blue line increases and the chances of observing double peaked emission decrease. 

This is also found in semi-analytical and hydrodynamical models of the EoR, which typically predict negligible transmission bluewards of the systemic velocity at $z>6$ \citep{Dijkstra2007,Laursen2011,Weinberger2018}. In the radiative transfer simulations from \cite{Laursen2011} the median IGM transmission at the blue peak is $>30$ \% only at $z<4.5$, while it is $<0.1$ \% at $z>5$ (similar to more recent simulations by \citealt{Weinberger2018}). Given that these models predict such low transmission bluewards of the systemic redshift, how is it possible that we observe a strong blue peak in the Ly$\alpha$ profile of COLA1 at $z=6.59$, a redshift where this is highly unexpected? Here, we propose three scenarios that facilitate the transmission of blue Ly$\alpha$ photons and that may explain COLA1's Ly$\alpha$ profile.

\subsection{Three scenarios to explain double peaked Ly$\alpha$ emission at $z=6.6$}
In Fig. $\ref{fig:igmillustration}$, we sketch three different physical scenarios that may explain the Ly$\alpha$ profile of COLA1. We first describe these models and then we propose observations that may test and differentiate between them. We note that the transmission, line-widths and offsets in Fig. $\ref{fig:igmillustration}$ are chosen for illustrative purposes. 

In Model 1, COLA1 is surrounded by a highly ionised bubble that is large enough for Ly$\alpha$ photons to redshift out of resonance due to the expansion of the Universe before encountering a relatively neutral IGM \citep[e.g.][]{Haiman2002,Hu2016}. The Ly$\alpha$ profile is therefore only determined by the ISM conditions of the galaxy. In Model 2 the bubble itself is smaller, but is embedded in a larger relatively ionised region. The Ly$\alpha$ profile here is mostly affected by the gas in the CGM and IGM. Model 3 attempts to explain COLA1's Ly$\alpha$ profile without invoking IGM attenuation, but rather by an infalling self-shielded cloud of neutral hydrogen. This neutral hydrogen cloud absorbs part of the emitted Ly$\alpha$ line, resulting in a double peaked spectrum.

\subsubsection{Model 1: Large, highly ionised bubble in a neutral IGM}  \label{sec:model1}
If COLA1 resides in a large, highly ionised region, it is possible that blue Ly$\alpha$ photons redshift out of the resonance wavelength due to the Hubble expansion prior to encountering a neutral IGM \citep[e.g.][]{MalhotraRhoads2006}. Depending on the size of this highly ionised region, it is possible that only (part of) the blue peak is attenuated \citep[e.g.][]{Haiman2002}. In the left panel of Fig. $\ref{fig:igmillustration}$, we illustrate this scenario where a double peaked Ly$\alpha$ line escapes from the ISM of the galaxy (resulting from resonant scattering effects, e.g. \citealt{Neufeld1990,GronkeDijkstra2017}), before the blue part of the line is attenuated by a sigmoid transmission function of the IGM. 
 
The Ly$\alpha$ photons in the blue peak of COLA1 need to redshift out of the resonance wavelength by at least $\gtrsim250$ km s$^{-1}$ (the maximum velocity at which we observe blue Ly$\alpha$ flux compared to line-centre, see Fig. $\ref{fig:cola1_lya}$) before encountering significant amounts of neutral hydrogen. This requires a large ionised region. Ignoring peculiar velocities of inflowing gas, we can calculate the required size as follows:
\begin{equation}
{\rm d_{prop}}= \frac{\Delta v}{H_{z=6.59}} {\rm Mpc.}
\end{equation}
Here, d$_{\rm prop}$ is the proper distance photons are required to travel, $\Delta v$ the required velocity offset and $H_{z=6.59}$ is the Hubble parameter at $z=6.59$, which is $H_{z=6.59}=803.85$ km s$^{-1}$ Mpc$^{-1}$ in our assumed cosmological model. Therefore, in order for Ly$\alpha$ photons to redshift by $\gtrsim250$ km s$^{-1}$ would require an ionised sightline/region of at least 0.3 pMpc; or $\gtrsim2.3$ cMpc. Simulations indicate that this bubble-size is similar to the characteristic bubble size for a global ionised fraction of $x_{\rm HI} \approx 50$ \% \citep[e.g.][]{Furlanetto2006,Lin2016}. As the ionised region needs to be {\it at least} 0.3 pMpc, and could be larger, this means the bubble size of COLA1 would correspond to a mean IGM $x_{\rm HI}<0.5$ at $z=6.6$. On the other hand, the bubble around COLA1 may also be an outlier as COLA1 is likely a relatively massive galaxy. This could imply a higher neutral fraction.

Can such an ionised region be explained using the observed properties of COLA1? Following \cite{Haiman2002},  we estimate the maximum proper radius of the ionised region around a galaxy (in the absence of neighbouring ionising sources and ignoring recombinations):
\begin{equation}
R_s = 2.5 (f_{\rm esc, LyC}\, Q_{ion} /10^{54} {\rm s}^{-1})^{1/3} (t_{\rm burst}/10^7 {\rm yr})^{1/3} (1+z)^{-1} \,{\rm Mpc}.
\end{equation}
Here, $R_s$ is the radius of the Str\"omgren sphere, $f_{\rm esc, LyC}$ the escape fraction of ionising photons, $Q_{\rm ion}$ the produced number of ionising photons per second and $t$ the age of the burst of star formation. As listed in Table $\ref{tab:IGMproperties}$, we estimate $f_{\rm esc, LyC}=0.15$ (based on the Ly$\alpha$ peak offset), $Q_{\rm ion} \approx 5-7\times10^{54}$ s$^{-1}$ (based on either the Ly$\alpha$ luminosity, or based on the UV luminosity, combined with a ionising photon production efficiency of $\xi_{ion}=10^{25.4}$ Hz erg$^{-1}$; \citealt{Bouwens2016b}) and $t_{\rm burst}=10^7$ yr (based on the high Ly$\alpha$ EW; e.g. \citealt{CharlotFall1993}) at $z=6.591$. This results in $R_{s, \rm max, Ly\alpha} = 0.33^{+0.07}_{-0.06}$ pMpc\footnote{We note that these uncertainties are the propagated errors corresponding to the UV and Ly$\alpha$ luminosities and ignore uncertainties in the age of the burst of star formation and the escape fraction.} and $R_{s, \rm max, UV}=0.29^{+0.03}_{-0.03}$ pMpc, corresponding to $\approx2.5$ cMpc. The maximum radius would marginally increase by 0.03 pMpc when correcting the UV luminosity for a (high) attenuation $A_{\rm UV}=0.45$.

Therefore, under basic assumptions, the star-formation in COLA1 may provide enough photons that can ionise a large enough region for allowing the blue peak to be observed up to $\approx-250$ km s$^{-1}$ from line-centre. However, there are important caveats that require attention. For example, our estimate conservatively assumes that the IGM does not affect Ly$\alpha$ photons red-wards of line-centre (as illustrated in the left panel of Fig. $\ref{fig:igmillustration}$) and the real required ionised region may have to be significantly larger. 

Moreover, the calculation so far also ignores peculiar velocities that typically blueshift Ly$\alpha$ photons with respect to neutral gas in the IGM \citep{Laursen2011}. Our calculation also ignores self-shielded neutral regions in the CGM around galaxies, that could be challenging to ionise by the galaxy itself and therefore may form a major source of opacity \citep[e.g.][]{Mesinger2015,Sadoun2017}. The recent simulations from \cite{Weinberger2018} suggest that these self-shielded regions may be more common for galaxies that reside in halos with M$_{\rm halo} \sim10^{11}$ M$_{\odot}$, for which the fraction of sight-lines that has a significant transmission on the blue side of the systemic redshift is consequently extremely low. Sight-lines with high blue-transmission do exist in a significant fraction of low mass halos (M$_{\rm halo} \sim10^{9}$ M$_{\odot}$). These simulations therefore suggest that COLA1 resides in a low mass halo, unless the ionisation from galaxies themselves have been under-estimated. 

Finally, as noted in \cite{Haiman2002}, the radius of the ionised sphere is over-estimated in case recombinations are important, for example when the clumping factor in the IGM is high ($C>10$) or the age of the star formation burst is $>10^8$ yr. It is thus unclear whether the ionised bubble can also be sustained, in particular when the star formation rate of COLA1 would decline or if the escape fraction would decrease. A solution would be if a quasar or (faint) neighbouring galaxies contribute to the local ionising budget \citep[e.g.][]{Kakiichi2018}. COLA1 is not located nearby a known quasar \citep{Banados2016}. However, COLA1 is at a relatively close separation to CR7, the most luminous LAE known at $z=6.6$. The projected distance on the sky of $34.17'$, which corresponds to a comoving distance of 77 Mpc (proper distance 10.1 Mpc) and the velocity difference is $\approx 350$ km s$^{-1}$. No (faint) neighbouring galaxies are known around COLA1 \citep{Bowler2014,Matthee2015}, although the sensitivity to neighbouring galaxies is limited (corresponding to SFR$\gtrsim25$ M$_{\odot}$ yr$^{-1}$) and deeper observations are required.

Our calculations differ somewhat from those in \cite{Hu2016}, who argue that a change in the Ly$\alpha$ properties is expected when the number of escaping ionising photons is $\approx10^{54}$ s$^{-1}$ (corresponding to $R_s \approx 0.3 $ pMpc and hence consistent with our assumptions above). While \cite{Hu2016} find that this is achieved with f$_{\rm esc, LyC} = 1$ \% in case of f$_{\rm esc, Ly\alpha} = 100$ \% and a Ly$\alpha$ luminosity $\approx2.5\times10^{43}$ erg s$^{-1}$, we find that this requires a higher escape fraction f$_{\rm esc, LyC} \approx 30$ \% at fixed luminosity. This can be mitigated to f$_{\rm esc, LyC} \approx 20$ \% in case of a slightly higher Ly$\alpha$ luminosity and lower Ly$\alpha$ escape fraction (see Table $\ref{tab:IGMproperties}$), indicating that a change in the Ly$\alpha$ line properties is still expected around COLA1's Ly$\alpha$ luminosity. 

\begin{table*}
\centering
\caption{Parameters used in calculations of the ionised bubble around COLA1 at $z=6.591$ (\S $\ref{sec:model1}$).} \label{tab:IGMproperties}
\begin{tabular}{lrr} \hline
\bf Property & \bf Value & \bf Motivation \\ \hline
$f_{\rm esc, LyC}$ & 15 \% & (Conservative) Ly$\alpha$ peak separation and blue-to-red flux ratio in local LyC leakers \\
$f_{\rm esc, Ly\alpha}$ & $60^{+20}_{-15}$ \% & Ly$\alpha$ EW calibration, tested at $z=0-2.6$ (\citealt{SobralMatthee2018}) \\
$Q_{\rm ion, Ly\alpha}$ &  $7^{+5}_{-3}\times10^{54}$ s$^{-1}$ & Ly$\alpha$ luminosity, $f_{\rm esc, LyC}$ and $f_{\rm esc, Ly\alpha}$ (\citealt{SobralMatthee2018}) \\
$Q_{\rm ion, UV}$ &  $5^{+1}_{-1}\times10^{54}$ s$^{-1}$ & UV luminosity (no dust correction) and $\xi_{ion}=10^{25.4}$ Hz erg$^{-1}$ \\
$t_{\rm burst}$ & $10^7$ yr & High Ly$\alpha$ EW (\citealt{CharlotFall1993}) \\
\hline
$R_{s, \rm required}$ & $>0.3$ pMpc & Velocity offset $\gtrsim250$ km s$^{-1}$ due to expansion of the Universe\\
$R_{s, \rm max, Ly\alpha}$ & $0.33^{+0.07}_{-0.06}$ pMpc & Eq. 2, assuming no recombinations and ionisation by the Ly$\alpha$ source of COLA1\\
$R_{s, \rm max, UV}$ & $0.29^{+0.03}_{-0.03}$ pMpc & Eq. 2, assuming no recombinations and ionisation by the UV source of COLA1\\ 
\hline
\end{tabular}
\end{table*}

\subsubsection{Model 2: Large mildly ionised region}
In our second scenario, COLA1 resides in a relatively large ionised region with a high transmission away from line-centre, while the galaxy itself is not able to ionise gas clouds in its direct vicinity, causing the low transmission at line-centre. This scenario is illustrated in the centre panel of Fig. $\ref{fig:igmillustration}$. Here, the double peaked emission does not emerge from the galaxy itself, but is caused by the environment of COLA1 that transmits flux at different velocities differentially. 

Here, the gas in the CGM of the galaxy still contains significant amounts of neutral hydrogen (e.g. due to self-shielding), while the gas at larger distances from the galaxy is mildly ionised and has a transmission of $\approx 50$ \% at $<-1000$ km s$^{-1}$ from the systemic redshift (similar to the simulated IGM properties at $z\approx4$ in \citealt{Laursen2011}). The double peaked emission originates from residual hydrogen in the vicinity of COLA1, absorbing flux around the systemic velocity. In this scenario, the intrinsic Ly$\alpha$ profile (that escapes the ISM) is broader than the observed red Ly$\alpha$ line, but the resulting blue and red lines have similar widths. While this scenario still requires a highly ionised region (a transmission of 50 \% corresponds to $\tau = 0.7$, or a mean neutral fraction of $\sim2\times10^{-6}$ at $z=6.6$), less ionising photons originating from COLA1 are required compared to Model 1.

We point out that double peaked Ly$\alpha$ profiles with low flux at line centre are the norm among Green Pea galaxies in the local Universe ($z\approx0.3$; e.g. \citealt{Henry2015,Yang2017}), emerging from scattering in the ISM. Therefore, it would be a remarkable coincidence if an emerging spectrum at $z=6.6$ would have a similar profile, but due to the effect of the IGM (and without scattering in the ISM).

\subsubsection{Model 3: Large velocity offset and intervening absorber}
The last scenario follows an alternative explanation which does not require a strong imprint from the IGM (either highly ionised, neutral, or partly ionised) in the vicinity of the emitting galaxy. Here, the emergent Ly$\alpha$ spectrum consists of a significantly redshifted broad peak (e.g. due to an outflow), such that line is not affected by H{\sc i} in the CGM or IGM. The right panel of Fig. $\ref{fig:igmillustration}$ shows that the double peaked profile originates from an absorbing system on top of this emission profile. This scenario is similar to the `cold flow' scenario that is used to explain the Ly$\alpha$ profiles of high-redshift radio galaxies \citep[e.g.][]{vanOjik1997}, but with the difference that the column density is much lower.

The absorbing system could be interpreted as an inflow \citep[e.g.][]{Matsuda2006}, or potentially a wall between two ionised regions (although we note that the IGM in model 3 can also be fully neutral), such as a filament that has not yet reionised \citep[e.g.][]{Finlator2009}. We explore the H{\sc i} column density that such an absorber would imply by fitting the line-profile with a voigt profile absorber on top of a gaussian emission line. The upper H{\sc i} column density limit is mostly determined by the width of the absorber, which implies a marginally self-shielded column density N$_{\rm HI} \lesssim 10^{17.7}$ cm$^{-2}$, blue shifted by $\approx80$ km s$^{-1}$ with respect to the peak of the emission. The intrinsic line-width FWHM is $380\pm20$ km s$^{-1}$.

How likely is such a scenario? In this model, both observed lines are redshifted with respect to the systemic redshift. This is not seen in $z=0-3$ galaxies with double peaked Ly$\alpha$ emission, where the systemic redshift typically lies between peaks \citep[e.g.][]{Kulas2012,Henry2015,Sobral2018b}. \cite{Vanzella2018} identify a luminous system at $z=4$ (M$_{1500}=-22.2$, L$_{\rm Ly\alpha} \approx 1.4\times10^{43}$ erg s$^{-1}$) that has four Ly$\alpha$ components, of which only one is redshifted by $\approx350$ km s$^{-1}$, while the others are either systemic or blueshifted. Multiple redshifted lines are seen in high-redshift radio galaxies \citep{vanOjik1997}, but these lines are typically much broader ($\rm FWHM\gtrsim1000$ km s$^{-1}$) and likely a consequence of large amounts of hydrogen present around the massive galaxies that harbour radio AGN. Therefore, unless particular conditions in the EoR are important, this scenario may be quite unlikely.

\subsubsection{Summary and predictions}
The sketched scenarios in Fig. $\ref{fig:igmillustration}$ may be over-simplified and reality may be a combination of different aspects of the models. Even though Models 2 and 3 may be unlikely, some aspects of these models may be combined with the general scenario in Model 1. For example, the absorber from Model 3 could reside in the highly ionised region from Model 1. Other complications would arise if the galaxy is undergoing a major merger, with multiple emission line regions at different relative velocities (although there are currently no indications for this scenario). How can observations discriminate between the different explanations and make progress?

As illustrated in Fig. $\ref{fig:igmillustration}$, it is clear that the key quantities that would allow to differentiate between the models are the intrinsic line-width and the systemic redshift (marked as dashed,
vertical lines in Fig. $\ref{fig:igmillustration}$). These properties can be measured from the H$\alpha$ or the H$\beta$ lines, accessible with the {\it James Webb Space Telescope}. The systemic redshift can alternatively also be measured using [C{\sc ii}] FIR emission with ALMA \citep[e.g.][]{Matthee2017ALMA}, which can also test whether a merger of multiple components is ongoing.
Furthermore, if COLA1 is surrounded by a large ionised region, additional sources of ionising photons are likely present. It is therefore expected that COLA1 resides in a larger scale over-density of faint galaxies, potentially observable using deep Ly$\alpha$ observations with VLT/MUSE and/or with ALMA observations that can identify [C{\sc ii}].

\subsection{Implications \& considerations for surveys}
Why have only a few double peaked LAEs been found at $z=5-7$ \citep[e.g.][]{Songaila2018}? Why do we find a double peaked LAE at a redshift where the IGM transmission towards the blue is likely very low, instead of around $z\approx5-6$? Could it be that more double-peaked LAEs have been found, but they have been discarded as [O{\sc ii}] interlopers? 

\subsubsection{Observational biases against identifying double peaked LAEs?}
As already noted in \cite{Dijkstra2007}, LAEs with a strong blue peak show a relatively low asymmetry (quantified by skewness) and could thus be classed as low-redshift interlopers in the case of low resolution spectroscopy. Moreover, the range of observed peak separations in double peaked LAEs ($\Delta {\rm v} \approx 200-400$ km s$^{-1}$; \citealt{Verhamme2017}) corresponds to observed wavelength differences $\Delta \lambda_{\rm obs} = (1+z_{\rm Ly\alpha})\times[0.8-1.6]$ {\AA}, similar to the observed wavelength difference of the [O{\sc ii}] doublet  $\Delta \lambda_{\rm obs, [\rm OII]} = (1+z_{\rm [OII]})\times2.79$ {\AA} at all observed wavelengths in the optical. Therefore, `single' line-identifications may be biased against confirming double peaked LAEs. This bias may particularly be important when spectral wavelength coverage is limited to the optical, but less important for $z_{\rm Ly\alpha}\lesssim4.8$, as potential interlopers of LAEs at these wavelengths will likely be detected in multiple lines (such as H$\beta$ and [O{\sc iii}], besides [O{\sc ii}]).

There are also other explanations of why the number of known double-peaked LAEs at $z=5-6$ may be low. One explanation is that the fraction of double-peaked LAEs increases strongly with Ly$\alpha$ luminosity (as discussed in \citealt{Hu2016}) and current samples at $z=5-6$ do not yet include enough luminous sources (as the majority of very wide-field surveys prioritised $z>6$; e.g. \citealt{Songaila2018}). As the typical Ly$\alpha$ luminosity increases with redshift, this explanation would indirectly imply that the LyC escape fraction increases with Ly$\alpha$ luminosity and with redshift (as L$^{\star}_{\rm Ly\alpha}$ increases; \citealt{Sobral2018}).

Alternatively, it may be possible that spectroscopic follow-up observations and/or publication efforts may be biased towards $z>6$ and double-peaked LAEs at $z=3-6$ simply have not yet been observed spectroscopically or published; or that the typical S/N of spectroscopic observations at $z>4$ is too low to identify similar blue peaks. The flux in the blue peak is a factor $\approx0.3$ lower than the brighter red component and can therefore only be detected at 3$\sigma$ significance if the red peak is detected at $\approx10\sigma$ significance, which is not always the case.

\subsubsection{A preference for observing narrow peak separations only in luminous LAEs at $z>5$?}
Multiple peaked Ly$\alpha$ profiles have regularly been observed in LAEs at $z=2-3$ \citep[e.g.][]{Rauch2008,Rauch2011,Kulas2012,Yamada2012,Trainor2015,Vanzella2016,RiveraThorsen2017}, but also in quasars and in high-redshift radio galaxies \citep[e.g.][]{vanOjik1997,MileyDeBreuck2008}. As COLA1 does not show evidence for AGN activity, we focus our comparison to LAEs at $z=2-3$ (combining data from \citealt{Yamada2012}, \citealt{Saez2015} and \citealt{Sobral2018b}) and Ly$\alpha$ analogues (i.e. Green Pea galaxies; GPs; \citealt{Yang2017}). The fraction of double peaked LAEs at $z=2-3$ is $\approx30-50$ \% \citep{Kulas2012,Yamada2012,Trainor2015}, who find typical peak separations of $\sim500-750$ km s$^{-1}$, although we note that the spectral resolution of these observations ($\approx150-200$ km s$^{-1}$) may not identify peaks with low velocity separations. The peak separation in COLA1 is smaller than the peak separation of the ultra-faint (M$_{1500}=-17$), lensed LAE at $z=3.12$ identified in \cite{Vanzella2016}, which is an extremely compact, highly ionising star-forming galaxy. 

\begin{figure}
\centering
	\includegraphics[width=9.0cm]{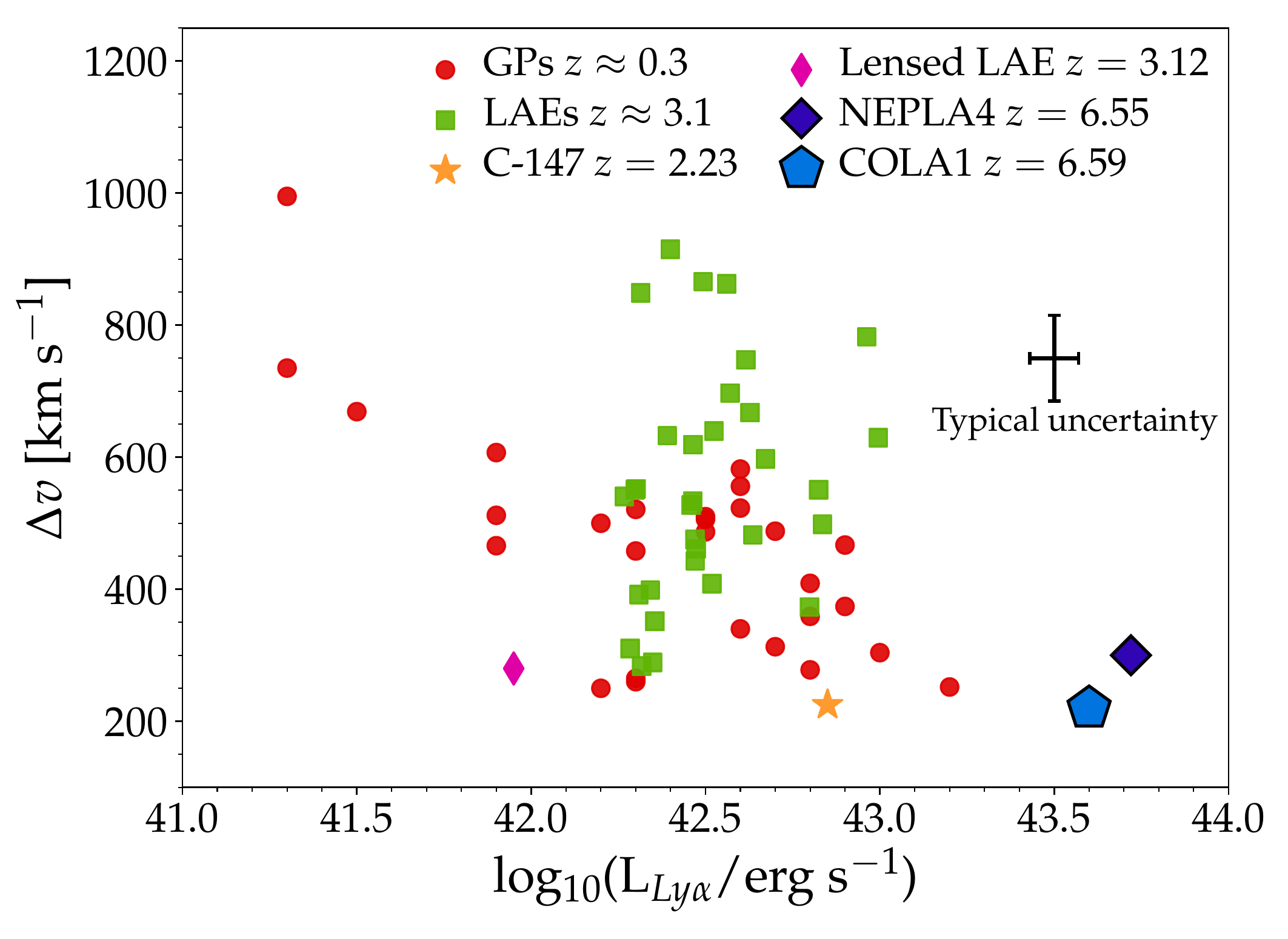} 
    \caption{Ly$\alpha$ peak separation versus the observed Ly$\alpha$ luminosity of a compilation of star-forming LAEs. COLA1 ({\it blue pentagon}) has the lowest peak separation known, while NEPLA4 ({\it dark blue diamond}; \citealt{Songaila2018}) is the most luminous LAE known with a double peak. While there is no clear relation between the peak separation and Ly$\alpha$ luminosity at $z\approx2-3$ ({\it orange star}: luminous LAE C-147 at $z=2.23$, that is shown in Fig. $\ref{fig:visspec}$; {\it green squares}: points based on published results from \citealt{Yamada2012} and \citealt{Saez2015}; {\it magenta diamond}: a low mass M$_{\rm star}<10^7$ M$_{\odot}$ compact LAE presented in \citealt{Vanzella2016}), a clear anti-correlation is seen between peak separation and luminosity for Ly$\alpha$ emitting Green Pea galaxies (GPs; {\it red points}, \citealt{Yang2017}). This anti-correlation may be a selection effect, as GPs are selected on compactness and strong [O{\sc iii}]/[O{\sc ii}] ratios that may trace density bounded H{\sc ii} regions, and not a pure Ly$\alpha$ selection.  }
    \label{fig:peaksep}
\end{figure}

As shown in Fig. $\ref{fig:peaksep}$, there is no clear relation between peak separation and Ly$\alpha$ luminosity for LAEs at $z\approx3$ (although the dynamic range is limited). On the other hand, a clear anti-correlation is seen between peak separation and luminosity for Ly$\alpha$ emitting GPs. COLA1, which is the most luminous LAE with the lowest known peak separation, seems to follow this relation. However, we note that this anti-correlation may be a selection effect, as GPs are selected on compactness and strong [O{\sc iii}]/[O{\sc ii}] ratios \citep{Cardamone2009}, that may trace density bounded H{\sc ii} regions. For example, GPs may all have a similar intrinsic Ly$\alpha$ luminosity, but because of a correlation between the Ly$\alpha$ escape fraction and the peak separation \citep{Yang2017} a strong anti-correlation is observed (Fig. $\ref{fig:peaksep}$).

If Ly$\alpha$ luminosity is anti-correlated with the peak separation at $z>6$, this may explain why double peaked Ly$\alpha$ profiles may only be observed in the most luminous LAEs. As the IGM transmission strongly decreases towards the blue of the systemic redshift \citep[e.g.][]{Laursen2011,Weinberger2018}, the blue lines of faint LAEs are more likely to be reduced. Luminous LAEs may however reside in larger ionised bubbles (because of a higher amount of ionising photons), facilitating the observability of a blue peak. This would for example imply that NEPLA4 \citep{Songaila2018}, resides in a larger ionised bubble than COLA1. \footnote{NEPLA4 has a larger peak separation than COLA1, inconsistent with the peak separation that an [O{\sc ii}] interloper would have.} On the other hand, simulations argue that the blue peak transmission may be higher for fainter galaxies, as there is less self-shielded H{\sc i} present in their CGM \citep[e.g.][]{Weinberger2018}. The low {\it observed} peak separation in COLA1 could also indicate that the CGM/IGM attenuated a significant amount of the flux at even bluer wavelengths. Hence, it may simply be expected that observed peak separations at $z>6$ are lower than at $z<5$. A crucial test would therefore be to map out the evolution of the relation between peak separation and Ly$\alpha$ luminosity from $z\approx5-7$ with high resolution spectroscopy, and explore how the blue-to-red flux ratio evolves with redshift.

The Ly$\alpha$ observability of luminous galaxies at $z\sim7$ appears to be boosted with respect to the observability of fainter galaxies \citep[e.g.][]{Stark2017}. While recent models with uniform ionising backgrounds reproduce this result \citep[e.g.][]{Mason2018Letter}, they can not yet explain the highest Ly$\alpha$ equivalent widths. We speculate that this could be mitigated in case the local ionising background is enhanced around bright galaxies (i.e. similar to Model 1 discussed here).

\section{Summary} \label{sec:summary}
We have presented a detailed analysis of the COLA1 galaxy, a double peaked LAE at $z=6.59$ \citep{Hu2016}, using deep multi-wavelength photometry and new deep, high resolution optical and near infrared spectroscopy with VLT/X-SHOOTER. Due to the opacity of the CGM and IGM for blue Ly$\alpha$ photons, detecting a blue peak at $z>6$ is highly unlikely. We therefore test whether the line profile is interpreted correctly as Ly$\alpha$ at $z=6.6$ and study the properties of COLA1 and their implications. Our results are summarised as follows:
 
\begin{itemize}

\item We rule out that the double peaked line is [O{\sc ii}] emission at $z=1.475$ based on 1) the shape of the line-profile: the asymmetric wing of the red line is not observed in the blue line and no flux is detected in the middle of the blue and red lines; 2) the extremely low blue-to-red flux ratio that is inconsistent with [O{\sc ii}] and 3) the non-detection of H$\alpha$ at $z=1.475$ ruling out even the lowest observed ratios with respect to [O{\sc ii}]. 

\item A tentative detection in the $B$ band is explained by a contribution from Ly$\alpha$ emission in a foreground LAE at $z=2.142$, confirmed through [O{\sc iii}]$_{5007}$ emission in addition to Ly$\alpha$.

\item We confirm that COLA1 is a double peaked LAE at $z_{\rm Ly\alpha, red}=6.593$ and summarise its properties in Table $\ref{tab:lyashell}$. The Ly$\alpha$ luminosity and EW are high (L$_{\rm Ly\alpha}=4.1\pm0.2\times10^{43}$ erg s$^{-1}$ and EW$_{0}=120^{+50}_{-40}$ {\AA}) and the Ly$\alpha$ emission is compact ($r_{50} = 0.33^{+0.07}_{-0.04}$ kpc based on {\it HST} imaging). 

\item The Ly$\alpha$ peak separation is $220\pm20$ km s$^{-1}$ and the flux of the blue peak is $0.31\pm0.03$ times the flux in the red peak. The Ly$\alpha$ lines are narrow (FWHM$=198\pm14$ km s$^{-1}$ and $150\pm18$ km s$^{-1}$ for the red and blue peak, respectively) and the full Ly$\alpha$ line-profile resembles the profile of a luminous LAE at $z=2.2$ with a prominent blue peak (Fig. $\ref{fig:visspec}$). Due to the narrowness of the Ly$\alpha$ line, COLA1 is unlikely to be powered by an AGN.

\item By modelling the Ly$\alpha$ profile with a five-parameter shell model, we find that the line-profile is characterised by a very low H{\sc i} column density (N$_{\rm HI} = 10^{17.0\pm0.3}$ cm$^{-2}$), indicating a non-zero escape of ionising photons. Based on correlations between the escape fraction and the peak separation in LyC leaking galaxies, we infer an escape fraction of $\approx15$\%, but potentially up to $\approx30$ \%. Other inferences imply escape fractions up to f$_{\rm esc, LyC} \sim 50$ \%. This means that we are witnessing a star-forming galaxy that is actively contributing to the reionisation of the Universe.

\item COLA1 has a high UV luminosity, M$_{1500}=-21.6\pm0.3$, implying a $\rm SFR\approx30$ M$_{\odot}$ yr$^{-1}$. The {\it Spitzer}/IRAC colours $[3.6]-[4.5]=-0.2\pm0.3$ imply relatively weak contribution from [O{\sc iii}]/H$\beta$ contributing to the $[3.6]$ band, implying a metallicity as low as $Z<10^{-3} (1/20\,Z_{\odot})$ or reduced strength of nebular emission lines due to escaping ionising photons. 

\item The detectability of blue Ly$\alpha$ emission from COLA1 implies a low Ly$\alpha$ optical depth out to $\approx-250$ km s$^{-1}$ of the galaxy, potentially due to a surrounding $\gtrsim0.3$pMpc (2.3 cMpc) ionised region.
Based on the observed properties of COLA1 and simple assumptions (detailed in Table $\ref{tab:IGMproperties}$), we find that enough ionising photons have escaped COLA1 to ionise a $\approx0.3$pMpc (2.3 cMpc) bubble. However, keeping this region ionised likely requires a contribution from faint neighbouring sources. No faint neighbouring galaxies are known, but COLA1 is likely in the same large scale over-density as CR7 with a separation of $\approx30'$ ($\approx10$ pMpc) and $\approx 350$ km s$^{-1}$.

\end{itemize}

As we have shown, the confirmation of COLA1 as a double peaked LAE at $z=6.59$ allows us to witness directly that galaxies contribute to the reionisation of the Universe. Several assumptions made here can be overcome with future observations of the intrinsic line-profile and systemic redshift of COLA1. In particular, future high resolution spectroscopic observations of samples of LAEs may reveal the dependence on double peaked Ly$\alpha$ profiles on luminosity and redshift, providing a new probe of the reionisation process.

\section*{Acknowledgements} 
JM acknowledges the award of a Huygens PhD fellowship from Leiden University. MG acknowledges support from NASA grant NNX17AK58G. APA, PhD::SPACE fellow, acknowledges support from the FCT through the fellowship PD/BD/52706/2014. Based on observations made with ESO Telescopes at the La Silla Paranal Observatory under programme IDs 294.A-5018, 098.A-0819, 099.A-0254 and 0100.A-0213. We are grateful for the excellent data-sets from the COSMOS and UltraVISTA survey teams. This research was supported by the Munich Institute for Astro- and Particle Physics (MIAPP) of the DFG cluster of excellence "Origin and Structure of the Universe". We thank the referee for their comments that improved the paper. We also thank Christoph Behrens, Len Cowie, Koki Kakiichi, Peter Laursen, Charlotte Mason, Eros Vanzella, Lewis Weinberger and Johannes Zabl for discussions. We have benefited from the public available programming language \texttt{Python}, including the \texttt{numpy, matplotlib, scipy} and \texttt{astropy} packages \citep{Hunter2007,Astropy2013}, the astronomical imaging tools \texttt{Swarp} \citep{Swarp2010} and \texttt{ds9} and the \texttt{Topcat} analysis tool \citep{Topcat}.

\bibliographystyle{aa}

\bibliography{COLA1.bib}

\end{document}